\newif\ifconfver
\confverfalse      

\ifconfver
     \documentclass[10pt,twocolumn,twoside]{IEEEtran}
\else
     \documentclass[12pt, draftclsnofoot, onecolumn]{IEEEtran}
\fi

\usepackage{calc,amsfonts,amssymb,amsmath,bm,url,color,theorem,cite,verbatim}
\usepackage{graphicx,subfigure}
\usepackage{epsfig,multirow,epstopdf}
\usepackage{algorithmic,algorithm}
\usepackage[T1]{fontenc}

\newlength{\twidth}
\ifconfver
\else
\fi

    \makeatletter
    \def\multilimits@{\bgroup
  \Let@
  \restore@math@cr
  \default@tag
 \baselineskip\fontdimen10 \scriptfont\tw@
 \advance\baselineskip\fontdimen12 \scriptfont\tw@
 \lineskip\thr@@\fontdimen8 \scriptfont\thr@@
 \lineskiplimit\lineskip
 \vbox\bgroup\ialign\bgroup\hfil$\m@th\scriptstyle{##}$\hfil\crcr}
    \def\Sb{_\multilimits@}
    \def\endSb{\crcr\egroup\egroup\egroup}
\makeatother

\newtheorem{lemma}{Lemma}
\newtheorem{thm}{Theorem}

\newtheorem{prop}{Proposition}

\definecolor{orange}{RGB}{255,107,0}
\definecolor{purple}{rgb}{0.627,0.125,0.941}

\newcommand\stroke[1]{\mathpalette\stroke@aux{#1}}
\def\stroke@aux#1#2{%
  \ooalign{%
    \hfil$#1-$\hfil\cr
    \hfil$#1#2$\hfil\cr
  }%
}

{\begin{list}{}%
    {\setlength{\rightmargin}{0pc}%
    \setlength{\leftmargin}{1pc} }} %
{\end{list}}

{\begin{list}{}%
    {\setlength{\rightmargin}{1pc}%
    \setlength{\labelsep}{0pc}
    \setlength{\leftmargin}{1pc} }} %
{\end{list}}

{\begin{list}{}%
    {\setlength{\rightmargin}{0pc}%
    \setlength{\leftmargin}{0pc} }} %
{\end{list}}

{\begin{list}{}{
    \settowidth{\labelwidth}{\mbox{\textnormal{#1}}}%
    \setlength{\leftmargin}{\labelwidth+\labelsep}}}%
{\end{list}}

\begin{document}

\title{A Stochastic Beamformed Amplify-and-Forward Scheme in a Multigroup Multicast {MIMO} Relay Network with Per-Antenna Power Constraints}
\ifconfver \else 
{
\linespread{1.1} \rm \fi

\author{
Sissi Xiaoxiao Wu, Qiang Li, Anthony Man-Cho So and Wing-Kin Ma \\
    \thanks{This work was supported in part by the Hong Kong Research Grant Council (RGC) General Research Fund (GRF) Project CUHK 416012, in part by The Chinese University of Hong Kong Direct Grant No.~4055009, and in part by the National Natural Science Foundation of China Grant No.~61401073.}
		    \thanks{Sissi Xiaoxiao Wu is the corresponding author. She and Anthony Man-Cho So are with the
Department of Systems Engineering and Engineering Management, The Chinese University of Hong
Kong, Shatin, N.T., Hong Kong S.A.R., China. E-mail: xxwu@ee.cuhk.edu.hk, manchoso@se.cuhk.edu.hk.}
\thanks{Qiang Li is with the School of Comm. Info. Eng., University of Electronic Science Technology of China, China. E-mail: lq@uestc.edu.cn.}
\thanks{Wing-Kin Ma is with the Department of Electronic Engineering, The Chinese University of Hong Kong, Shatin, N.T., Hong Kong S.A.R., China. E-mail: wkma@ieee.org.}
}
}

\maketitle


\begin{abstract}
In this paper, we consider a two-hop one-way relay network for multigroup multicast transmission between long-distance users, in which the relay is equipped with multiple antennas, while the transmitters and receivers are all with a single antenna.  Assuming that perfect channel state information is available, we study amplify-and-forward (AF) schemes that aim at optimizing the max-min-fair (MMF) rate.  We begin by considering the classic beamformed AF (BF-AF) scheme, whose corresponding MMF design problem can be formulated as a rank-constrained fractional semidefinite program (SDP).  We show that the gap between the BF-AF rate and the SDR rate associated with an optimal SDP solution is sensitive to the number of users as well as the number of power constraints in the relay system.  This reveals that the BF-AF scheme may not be well suited for large-scale  systems.  We therefore propose the stochastic beamformed AF (SBF-AF) schemes, which differ from the BF-AF scheme in that time-varying AF weights are used. 
We prove that the MMF rates of the proposed SBF-AF schemes are at most $0.8317$ bits/s/Hz less than the SDR rate, irrespective of the number of users or power constraints.  Thus, SBF-AF can outperform  BF-AF especially in large-scale systems. Finally, we present numerical results to demonstrate the viability of our proposed schemes.
\ifconfver  \else
\\[.5\baselineskip]
\noindent
\end{abstract}

%

%
%

\section{Introduction}\label{sec:intro}

\begin{figure}[h]
\centering
  \includegraphics[width=0.36\textwidth]{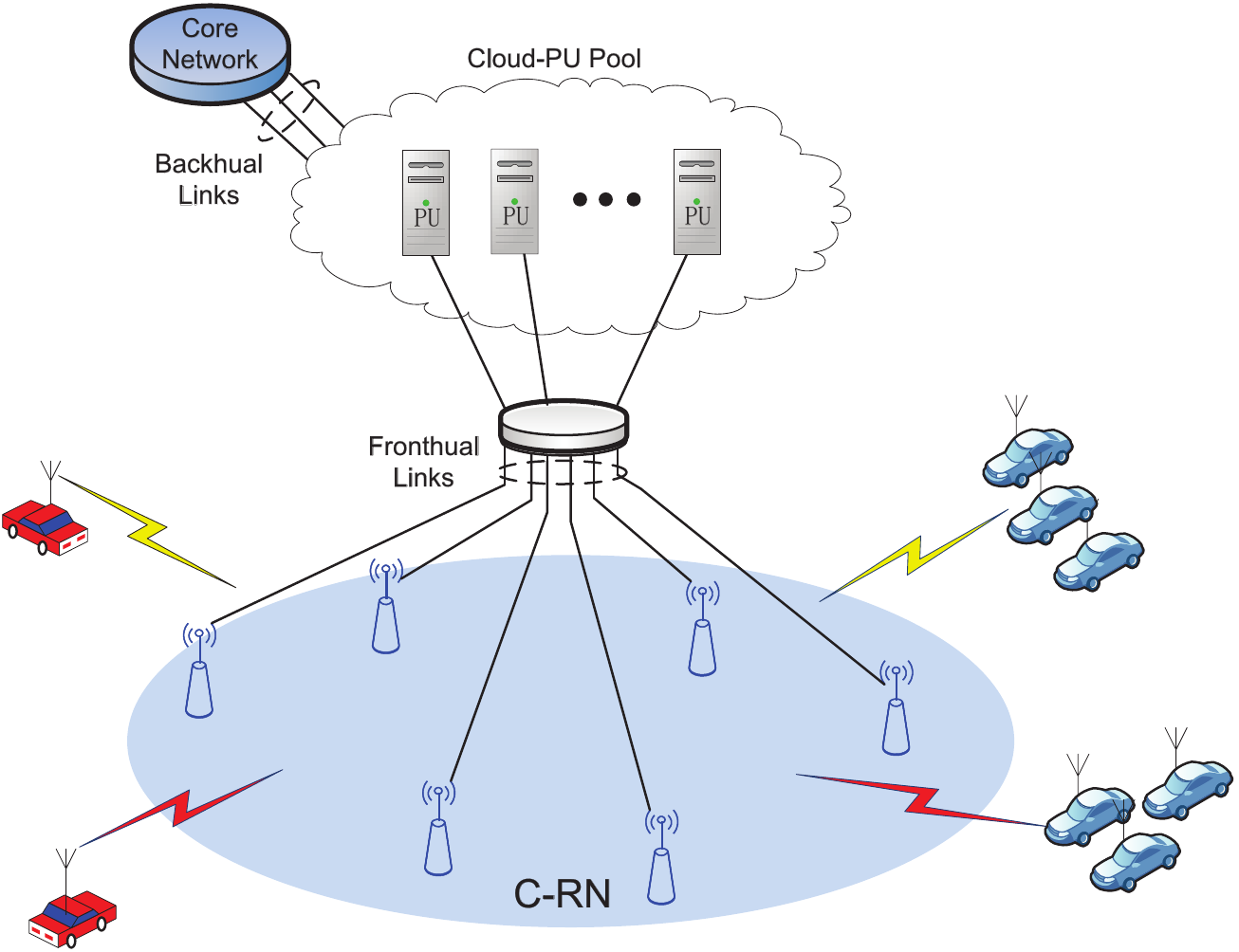}
\caption{An example of a cloud relay network.}
\label{fig:cloud_relay}
\end{figure}

It is well known that path loss, shadowing, and multi-path fading can cause a severe degradation of the channel between long-distance users. To overcome these effects, a popular approach is to employ relay nodes to amplify the signals of the transmitters and forward them to the receivers.  Besides supporting applications such as military communications and device-to-device (D2D) communications, where users are usually limited by power or apparatus, such an approach has also found its role in 5G broadband applications.  Indeed, there is a new trend of employing fronthaul-backhaul links to coordinate relay nodes to form a big MIMO relay station. For example, the studies of C-RAN \cite{wpcran,shi2013group,cranfronthaul,wucrn_r2} have led to the so-called \emph{cloud relay network (C-RN)} in \cite{wucrn_r2} (see Figure \ref{fig:cloud_relay}), where the channel state information (CSI) is perfectly known and fully shared, while data information is partially or fully shared within the cloud processing unit (PU) pool.{{\footnote{In practice, the limited capacity of the fronthaul and backhaul links of C-RN is also an important issue. Here, for simplicity, we do not impose any specific constraint on the link capacity and focus on the AF relaying design.}}} It is easy to see that if CSI and data information in the C-RN are both fully shared, then we are actually dealing with an MIMO relay network. This motivates us to study the design of amplify-and-forward (AF) schemes for such kind of networks.\footnote{The relays can also decode-and-forward (DF) the received signals, but this is beyond the scope of this paper.} 

In this paper, we consider a typical two-hop one-way MIMO relay network, where there is no direct link between far-apart transmitters and receivers, and reliable information delivery is facilitated by relays. We assume that the transmitters and receivers are all equipped with a single antenna, and that the CSI is perfectly known in the network. Our goal is to design the AF relay schemes so as to achieve good system performance. In the literature on MIMO relay networks, there are different formulations of such problem; see, e.g.,~~\cite{chalise2007mimo,chae2008mimo, rong2009unified, sanguinetti2012tutorial, Non_Regenerative_MIMO,Heath_MIMOrelay_08,Jnl:Relay_Chalise_09,Linear_MMSE,Precoder_Design,jimenez2012non,khandaker2012joint,MCY12,taowang12,chamaeh13,choi2014transceiver,dual_hop_MISO_relay2014}. Here, we focus on the multigroup multicast scenario and aim at optimizing the max-min-fair (MMF) rate. Towards that end, a classic approach is to adopt the beamformed AF (BF-AF) scheme~\cite{Jnl:Relay_Chalise_09}.  The MMF design problem corresponding to the BF-AF scheme can be formulated as a fractional quadratically-constrained quadratic program (QCQP), which is NP-hard in general~\cite{MulticastSidiropoulos06,Jnl:Karipidis_MM_2008}.  Nevertheless, the fractional QCQP is known to be equivalent to a rank-one constrained fractional semidefinite program (SDP), which can be tackled using the semidefinite relaxation (SDR) technique~\cite{Jnl:MagzineMaLUO}.  Roughly speaking, the SDR technique involves first computing an optimal solution to the fractional SDP without the rank constraint (which can be done efficiently).  Then, using a Gaussian randomization procedure, the optimal solution is converted into a rank-one solution, from which a feasible BF-AF solution can be extracted~\cite{chang2008approximation,jimulti13}. 
A natural question here is to quantify the gap between the MMF rate associated with the SDR-based BF-AF solution (which we call the \emph{BF-AF rate}) and the MMF rate associated with an optimal fractional SDP solution (which we call the \emph{SDR rate}).  Building upon the results in~\cite{chang2008approximation,jimulti13}, our first contribution is to show that in the worst-case, the gap is on the order of $\log M+\log\log L$ bits/s/Hz, where $M$ is the number of users in the MIMO relay network and $L$ is the number of power constraints on the relay antennas.  One immediate consequence of this result is that the BF-AF scheme may not be well suited for large-scale MIMO relay systems, where there are either many users or many power constraints.

The potentially large gap between the BF-AF rate and the SDR rate can be attributed to the mismatch between the rank of the SDR-based BF-AF solution (which is equal to one) and that of the optimal fractional SDP solution.  To improve the rate performance, one possibility is to design an AF relay scheme that can somehow utilize the information contained in the possibly high-rank optimal fractional SDP solution.  This motivates our second and main contribution of the paper, which is the design and analysis of stochastic BF-AF (SBF-AF) schemes for MIMO relay networks.  The key idea behind these schemes is to adopt time-varying random AF weights to simulate ``high-rank'' BF-AF.  This is achieved by choosing the distribution of the AF weights so that their covariance matrix is exactly equal to the optimal fractional SDP solution.  In this paper, we propose two SBF-AF schemes, which correspond to using the Gaussian and elliptic distributions to generate the AF weights, respectively.  Under some mild assumptions, we show that the MMF rates of the proposed SBF-AF schemes (which we call the \emph{SBF-AF rates}) are at most $0.8317$ bits/s/Hz less than the SDR rate.  Note that this bound is independent of the number of users or power constraints, which suggests that our proposed SBF-AF schemes can have a significant performance gain over the SDR-based BF-AF scheme, especially in large-scale MIMO relay systems.  As we shall see in Section~\ref{sec:sim}, such a claim is corroborated by our numerical results. 
{Moreover, the implementation of the SBF-AF schemes does not require the Gaussian randomization procedure.  Instead, it only requires the nodes in the network to have knowledge of a pre-specified random seed and then use it to perform beamformer generation and coherent detection (more implementation details are provided in Section III.C). Thus, the proposed SBF-AF schemes can reduce the computational complexity in the computing center of the network. We remark that some efficient heuristics have recently been proposed for finding a high-quality solution to a fractional QCQP; see, e.g., \cite{tran2014conic,christopoulos2015multicast,mehanna2015feasible,Gopalakrishnan15}. However, the fast convergence of these heuristics highly depends on a good initialization (such as the Gaussian randomization solution).  Moreover, there is no theoretical guarantee on the quality of the solutions found by these heuristics. By contrast, our proposed SBF-AF schemes enjoy strong theoretical properties.
}

The idea of stochastic beamforming (SBF)---i.e., using time-varying random beamformers to simulate ``high-rank'' beamforming---was first proposed in~\cite{MainPaper} for the single-group multicast scenario, where SBF is proven, both theoretically and numerically, to outperform transmit beamforming in terms of the multicast rate~\cite{MainPaper,wurobust}.  Our current work extends the works~\cite{MainPaper,wurobust} in two ways.  From the design perspective, we are the first to introduce SBF schemes in relay networks and expand their scope to cover the multigroup multicast scenario.  From the theoretical perspective, the rate performance analysis we conduct for the proposed SBF schemes is more involved than those in~\cite{MainPaper,wurobust}, as it needs to account for the interference in the system.  It should also be noted that the problem considered in this paper, namely beamformer design for multi-user to multi-user multigroup multicasting in MIMO relay networks, has not been well addressed in the literature.  Indeed, existing works on MIMO relay transceiver design mainly focus on the point-to-point~\cite{rong2009unified, sanguinetti2012tutorial, MCY12,dual_hop_MISO_relay2014,Non_Regenerative_MIMO,Linear_MMSE,Precoder_Design}, single-user to multi-user~\cite{jimenez2012non}, multi-user to single-user~\cite{khandaker2012joint}, and multi-user to multi-user unicast~\cite{choi2014transceiver,Heath_MIMOrelay_08,chalise2007mimo,Jnl:Relay_Chalise_09} and multicast~\cite{KR14} scenarios.  Although the work~\cite{BPG12} studies beamformer design in a multigroup multicast relay network, it only considers BF-AF schemes for single-antenna relays, whereas our focus is on SBF-AF schemes for a multi-antenna relay. {Moreover, it is worth mentioning that the same SBF technique developed in this paper is also applicable to multigroup multicasting in a standard MISO downlink scenario.} 



The paper is organized as follows. In Section~\ref{sec:BF}, we first introduce the system model of the MIMO relay network.  Then, we review the SDR-based BF-AF scheme and analyze its rate performance. Next, in Section~\ref{sec:sbf}, we develop the SBF-AF framework and analyze the rate performance of two SBF-AF schemes.  In Section~\ref{sec:DRN}, we discuss how the SBF-AF framework can be applied to a distributed relay network.  Then, we present numerical results on the performance of different AF schemes in Section~\ref{sec:sim}.  Finally, we conclude the paper in Section~\ref{sec:conclusions}.

Our notation is standard:
$\mathbb{R}^N$ and $\mathbb{C}^N$ are the sets of real and complex $N$-dimensional vectors, respectively;
$\mathbb{R}_+^N$ is the set of real $N$-dimensional non-negative vectors;
$\mathbb{H}_+^{N}$  is the set of $N \times N$ Hermitian positive semidefinite matrices;
$\| \cdot \|$ is the vector Euclidean norm;
${\bm A} \bullet {\bm B}$, ${\bm A} \otimes  {\bm B}$, and $\bm{A} \odot \bm{B}$ denote the inner product, Kronecker product, and Hadamard product between matrices ${\bm A}$ and ${\bm B}$, respectively;
${\rm rank}({\bm X})$, ${\rm \lambda}_{\rm max}({\bm X})$, and ${\rm \lambda}_{\rm min}^+({\bm X})$
stand for the rank, the largest eigenvalue, and the smallest non-zero eigenvalue of the matrix ${\bm X}$, respectively;
${\rm vec}({\bm A})$ is the vectorization of the matrix ${\bm A}$; 
${\rm Diag}(\bm{v})$ is the diagonal matrix with the vector $\bm{v}$ on the diagonal;
${\bm e}_i$ is the vector whose $i$th entry is 1 and the remaining entries are 0;
${\bm I}_r$ denotes the $r$-by-$r$ identity matrix;
$\mathbb{E}_{\bm{w}\sim\mathcal{D}}[ \cdot ]$ is the expectation operator with respect to the distribution $\mathcal{D}$ of the random vector $\bm{w}$;
$\mathcal{CN}({\bm 0},{\bm X})$ denotes the circularly symmetric complex Gaussian distribution  with mean vector ${\bm 0}$ and covariance matrix ${\bm X}$. 



\section{Problem Formulation and the SDR-based BF-AF Scheme}\label{sec:BF}

\subsection{System Model of the One-Way Relay Network} \label{subsec:MIMO}

\begin{figure}[htb]
\begin{center}
\includegraphics[width = 0.36\textwidth]{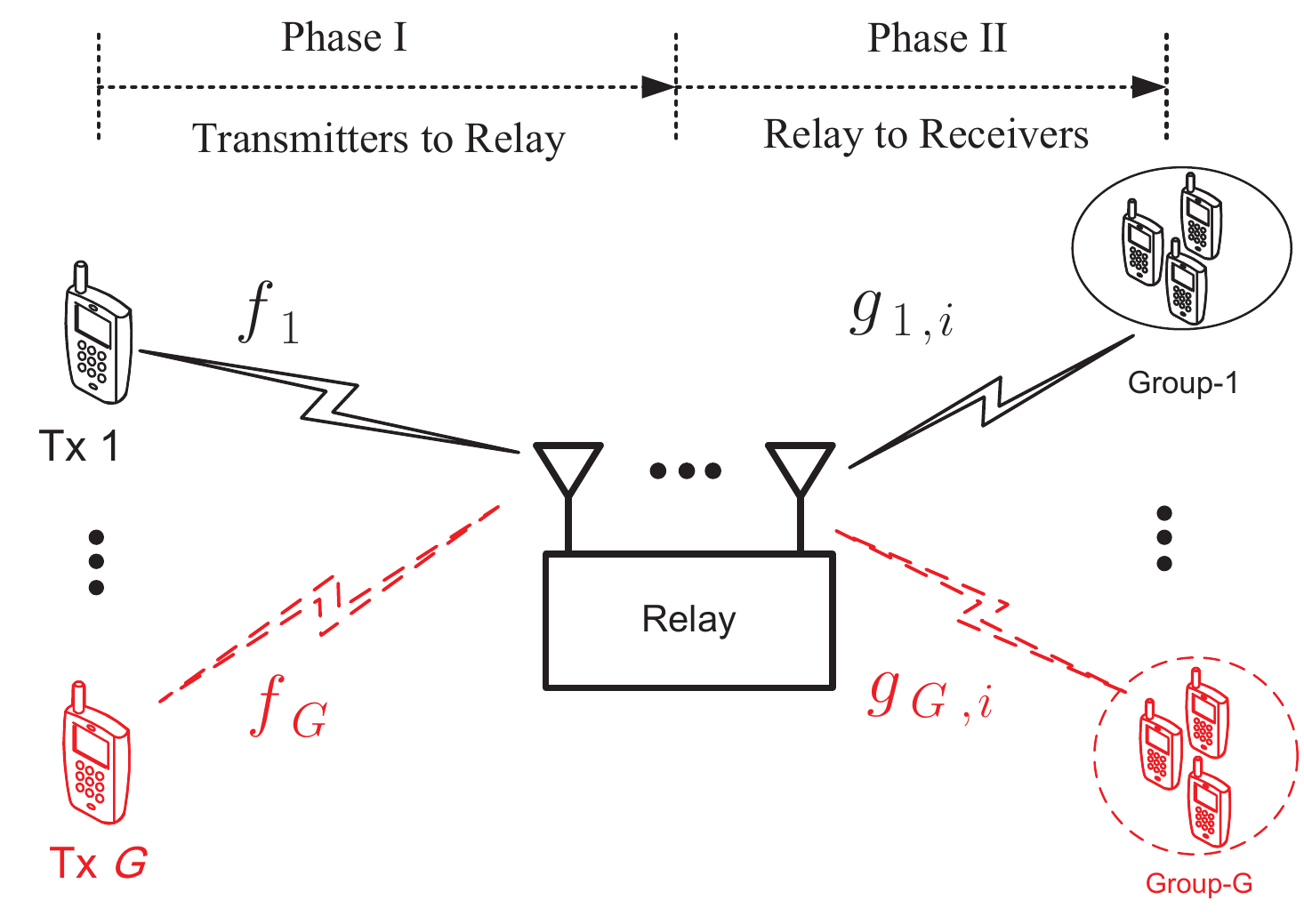}
\end{center}
\caption{The one-way MIMO relay multigroup multicasting model.}
\label{one_way_MIMO_relay}
\end{figure}

We consider multigroup multicast information delivery in an MIMO relay network as depicted in Figure \ref{one_way_MIMO_relay}. In the network, $G$ single-antenna transmitters send $G$ independent information streams to $G$ groups of single-antenna receivers (henceforth referred to as users).  Users in the same group request the same information, while users in different groups request different information.  Let $m_k$ denote the number of users in the $k$th group (where $k=1,\ldots,G$) and $M=\sum_{k=1}^G m_k$ denote the total number of users in the network.  We assume that there is no direct link between the transmitters and receivers, and reliable information delivery is enabled by the MIMO relay, which AF the signals received from the transmitters to the receivers. {We assume that the relay is equipped with $L$ antennas. Moreover, all the channels are quasi-static.}
Under this setting, the information delivery process consists of the following two phases:

\noindent
1) Phase I: \emph{Transmitters send information to relay.} The receive model of the transmitters-to-relay link is given by
\begin{equation}\label{rt}
\vspace{-0.1cm}
{\bm r}(t) = \sum_{j=1}^G{\bm f}_js_j(t) + {\bm n}(t),
\vspace{-0.1cm}
\end{equation}
where ${\bm r}(t) = \left[ \, r^1(t),\ldots,r^\ell(t),\ldots,r^L(t)\,\right]^T$ with ${r}^\ell(t) = \sum_{j=1}^G{f}_j^\ell s_j(t)+ {n}^\ell(t)$
being the received signal at the $\ell$th antenna of the MIMO relay; $s_j(t)$ is the common information designated for group~$j$ with $\mathbb{E}[|s_j(t)|^2]=P_j$, and $P_j$ is the transmit power at transmitter $j$; ${\bm f}_j= \left[ \, f_j^1,\ldots,f_j^\ell,\ldots,f_j^L\,\right]^T$ with $f_j^\ell$ being the channel from transmitter $j$ to the $\ell$th antenna of the MIMO relay;  ${\bm n}(t) = \left[ \, {n}^1(t),\ldots,{n}^\ell(t),\ldots,{n}^L(t)\, \right]^T$ with ${n}^\ell(t)$ being the mean zero, variance $\sigma_\ell^2$ Gaussian noise at the $\ell$th antenna of the relay.

\noindent
2) Phase II: \emph{Relay processes the received signals and forwards them to receivers.} A popular AF scheme in the literature is the BF-AF scheme~\cite{Jnl:Relay_Chalise_09}, which can be expressed as
\begin{equation} \label{eq:xt_v}
{\bm x}(t) = {\bm V} {\bm r}(t),
\end{equation}
where ${\bm V}$ is the AF weighting matrix.  The received signal of user $i$ in group $k$ is then given by
\begin{align}\label{yt}
y_{k,i}(t) &= {\bm g}_{k,i}^H{\bm x}(t) + v_{k,i}(t) \\\notag
&= \underbrace{{\bm g}_{k,i}^H{\bm V}{\bm f}_ks_k(t)}_{{\rm desired~signal}} \\\notag
&\quad+ \underbrace{{\bm g}_{k,i}^H{\bm V}\left(\sum_{m \neq k}{\bm f}_ms_m(t)\right)+ {\bm g}_{k,i}^H{\bm V}{\bm n}(t) + v_{k,i}(t)}_{{\rm interference~and~noise}},
\end{align}
where ${\bm g}_{k,i} = \left[ \, g_{k,i}^1,\ldots,g_{k,i}^\ell,\ldots,g_{k,i}^L\, \right]^T $ with $g_{k,i}^\ell$ being the channel from the $\ell$th antenna of the relay to user $i$ in group $k$; $v_{k,i}(t)$ is the Gaussian noise at user $i$ in group $k$ with mean zero and variance $\sigma_{k,i}^2$.  Under the above setting, the signal-to-noise-and-interference ratio (SINR) of user $i$ in group $k$ can be expressed as
\begin{align}\label{gamma}
  \frac{P_k\left|{\bm g}_{k,i}^H{\bm V}{\bm f}_k\right|^2}{\displaystyle\sum_{m \neq k}P_m\left|{\bm g}_{k,i}^H{\bm V}{\bm f}_m\right|^2+{\bm g}_{k,i}^H{\bm V}{\bm \Sigma}_L{\bm V}^H {\bm g}_{k,i}+ \sigma_{k,i}^2},
\end{align}
where ${\bm \Sigma}_L = {\rm Diag}(\sigma_1^2,\ldots,\sigma_L^2)$.

In designing the AF weighting matrix ${\bm V}$ for the MIMO relay, we consider two types of power constraints.  The first is the total power constraint on the relay.  Using~\eqref{eq:xt_v}, such a constraint can be formulated as
\begin{equation}\label{eq_total_power}
\mathbb{E}[\|{\bm x}(t)\|^2]={\rm Tr} \left({\bm V} \left( \sum_{j=1}^{G} P_j{\bm f}_j{\bm f}_j^H + {\bm \Sigma}_L \right){\bm V}^H \right) \le \bar{P}_0,
\end{equation}
where $\bar{P}_0>0$ is a given total power threshold.  The second is the per-antenna power constraints on the relay, which commonly arise in physical implementations of multi-antenna systems; see, e.g.,~\cite{yu2007transmitter,christopoulos2014weighted}.  These constraints can be formulated as
\begin{equation} \label{eq_per-power}
{\bm e}_{\ell}^H\underbrace{{\bm V} \left( \sum_{j=1}^{G} P_j{\bm f}_j{\bm f}_j^H+ {\bm \Sigma}_L \right) {\bm V}^H }_{\mathbb{E}[{\bm x}(t){\bm x}^H(t)]}{\bm e}_{\ell} \le \bar{P}_\ell, \quad \ell=1,\ldots,L,
\end{equation}
where $\bar{P}_\ell > 0$ is a given power threshold for the $\ell$th antenna of the relay.

By letting ${\bm w} = {\rm vec}({\bm V}) \in {\mathbb C}^{L^2}$ and using the identity
$$
{\rm Tr}\left({\bm A}^H{\bm B}{\bm C}{\bm D}\right)= {\rm vec}({\bm A})^H \left( {\bm D}^T \otimes {\bm B} \right) {\rm vec}({\bm C}),
$$
which is valid for arbitrary complex matrices ${\bm A}, {\bm B}, {\bm C}, {\bm D}$ of appropriate dimensions, we can express~\eqref{gamma} as
$$ \gamma_{k,i}({\bm w}{\bm w}^H) = \frac{{\bm w}^H {\bm A}_{k,i}{\bm w}}{{\bm w}^H {\bm C}_{k,i}{\bm w}+1}, $$
where
\begin{align}
{\bm A}_{k,i} &= P_k({\bm f}_k^* \otimes {\bm g}_{k,i})({\bm f}_k^* \otimes {\bm g}_{k,i})^H/\sigma_{k,i}^2, \label{ak_v} \\ 
{\bm C}_{k,i} & = \sum_{m \neq k}P_m({\bm f}_m^* \otimes {\bm g}_{k,i})({\bm f}_m^* \otimes {\bm g}_{k,i})^H/\sigma_{k,i}^2 \nonumber \\
&\quad+ {\bm \Sigma}_L \otimes \left( {\bm g}_{k,i}{\bm g}_{k,i}^H \right)/\sigma_{k,i}^2. \label{ck_v}
\end{align}
Similarly, we can rewrite constraints~\eqref{eq_total_power} and~\eqref{eq_per-power} as
\begin{equation} \label{eq:const-cpt}
{\bm w}^H {\bm D}_\ell{\bm w}\le \bar{P}_\ell, \quad \ell=0,1,\ldots,L,
\end{equation}
where 
\begin{align}
{\bm D}_0 &= \left( \sum_{j=1}^G P_j{\bm f}_j^*({\bm f}_j^*)^H + {\bm \Sigma}_L \right) \otimes {\bm I}_L, \label{d_v0} \\
{\bm D}_\ell &= \left( \sum_{j=1}^G P_j{\bm f}_j^*({\bm f}_j^*)^H + {\bm \Sigma}_L \right) \otimes \left( {\bm e}_{\ell}{\bm e}_{\ell}^H \right), \quad \ell=1,\ldots,L. \label{d_v}
\end{align}

\subsection{An SDR-Based MMF Formulation} \label{subsec:SDR}

Assuming that the CSI (i.e., ${\bm f}_k$ and ${\bm g}_{k,i}$) is perfectly known, we can now formulate the MMF design problem corresponding to the BF-AF scheme as
\begin{align*}
({\rm BF}) \quad {\bm w}^\star &= \displaystyle \arg \max_{{\bm w} \in \mathbb{C}^{L^2}}\displaystyle\min_{k=1,\ldots,G \atop i=1,\ldots,m_k}  {\gamma}_{k,i}({\bm w}{\bm w}^H) \\ 
&\quad\,\, \text{subject to} \,\,\, \eqref{eq:const-cpt}. 
\end{align*}
Problem (BF) is an instance of a fractional QCQP, which is NP-hard in general~\cite{MulticastSidiropoulos06,Jnl:Karipidis_MM_2008}.  Nevertheless, it can be tackled by the SDR technique~\cite{Jnl:MagzineMaLUO}.  
Specifically, upon observing that
\begin{equation} \label{eq:rank-eqv}
{\bm W} = {\bm w}{\bm w}^H \,\Longleftrightarrow\, {\bm W} \succeq {\bm 0}, \, \mbox{rank}({\bm W}) \le 1,
\end{equation}
we can relax Problem (BF) to the following fractional SDP:
\begin{align}\notag
{\rm (SDR)} \quad {\bm W}^\star &= \displaystyle \arg\max_{{\bm W} \in \mathbb{H}_+^{L^2}}\gamma(\bm{W}) \\ \label{contsdr}
\text{subject to} &\quad {\bm D}_\ell \bullet {\bm W} \le \bar{P}_{\ell}, \quad \ell=0,1,\ldots,L. 
\end{align}
Here, we define
\begin{equation}\label{eq:gamma}
\gamma(\bm{W}) =  \min_{k=1,\ldots,G \atop i=1,\ldots,m_k} \frac{{\bm A}_{k,i}\bullet {\bm W}} {{\bm C}_{k,i}\bullet {\bm W}+1}.
\end{equation}
{It is well known that (SDR) can be rewritten as 
\begin{align*}
&\displaystyle \max_{{\bm W} \in \mathbb{H}_+^{L^2},\,t}\quad t \\ 
\text{subject to} & \quad {\gamma}_{k,i}(\bm{W}) \ge t, \quad k=1,\ldots,G, i=1,\ldots,m_k, \\
& \quad \eqref{contsdr} ~{\rm is\,\,satisfied},
\end{align*}
whose solutions are in correspondence with those to the following power minimization problem~\cite{Jnl:Karipidis_MM_2008}:
\begin{align}\label{pm}
&\displaystyle \min_{{\bm W} \in \mathbb{H}_+^{L^2}}\quad {\bm D}_0 \bullet {\bm W} \\\notag
\text{subject to} & \quad {\gamma}_{k,i}(\bm{W}) \ge \gamma, \quad k=1.\ldots,G, i=1,\ldots,m_k, \\\notag
&\quad {\bm D}_\ell \bullet {\bm W} \le \bar{P}_{\ell}, \quad \ell=1,\ldots,L. 
\end{align}
Thus, the optimal value of Problem (SDR) can be approximated to arbitrary accuracy efficiently by performing a bisection search on $\gamma$, {where each iteration of the search involves solving the SDP~\eqref{pm} (see~\cite{christopoulos2015multicast,Jnl:Karipidis_MM_2008} for details).}
If $\mbox{rank}({\bm W}^\star) \le 1$, then by~\eqref{eq:rank-eqv}, we have ${\bm W}^\star={\bm w}^\star({\bm w}^\star)^H$ for some ${\bm w}^\star \in {\mathbb C}^{L^2}$.  Moreover, ${\bm w}^\star$ is optimal for (BF).
On the other hand, if ${\rm rank}(\bm{W}^\star) > 1$, then by applying a Gaussian randomization procedure (Algorithm~\ref{alg:0}; cf.~\cite{chang2008approximation,jimulti13}), we can generate a rank-one feasible solution $\widehat{\bm W}$ to (SDR) and extract from it a feasible but generally sub-optimal solution $\widehat{\bm w}$ to (BF).

Now, a fundamental issue is to quantify the quality loss of the solution $\widehat{\bm w}$ generated by Algorithm~\ref{alg:0}.  We shall tackle this issue from an achievable rate perspective and bound the achievable rate gap between the approximate solution $\widehat{\bm w}$ and the optimal solution ${\bm w}^\star$ to (BF).  To begin, let
$$
r_{\sf BF} = \log\left( 1+\gamma\left( \widehat{\bm w} \widehat{\bm w}^H \right) \right)
$$
be the BF-AF rate associated with the approximate solution $\widehat{\bm w}$.  Furthermore, let
$$
r_{\sf SDR} = \log\left( 1+\gamma(\bm{W}^\star) \right)
$$
be the SDR rate associated with an optimal solution ${\bm W}^\star$ to (SDR).  Since $\gamma\left( {\bm W}^\star \right) \ge \gamma\left( {\bm w}^\star({\bm w}^\star)^H \right) \ge \gamma\left( \widehat{\bm w} \widehat{\bm w}^H \right)$, we clearly have $r_{\sf SDR} \geq r_{\sf BF}$.  
{The following theorem shows that a reverse inequality (approximately) holds, which characterizes the quality of the solution return by Algorithm 1. }
\begin{thm} \quad\label{prop:1} 
{Let $M\ge1$ be the total number of users in the relay network and $L\ge2$ be the number of relay antennas in Problem (BF).\footnote{Here, we assume that $L \ge 2$, since we have ${\bm D}_0={\bm D}_1$ when $L=1$ in Problem~(BF).} Then, the following hold:
}
\begin{itemize}
\item[(a)] When $M+L \le 3$, an optimal solution ${\bm W}^\star$ to (SDR) with ${\rm rank}({\bm{W}^\star}) \le 1$ can be found efficiently.  Consequently, the solution $\widehat{\bm w}$ returned by Algorithm~\ref{alg:0} satisfies $r_{\sf BF} = r_{\sf SDR}$.

\item[(b)] When $M+L > 3$, the solution $\widehat{\bm w}$ returned by Algorithm~\ref{alg:0} will satisfy
\begin{equation} \label{eq:gap-bd}
r_{\sf SDR} - r_{\sf BF} \le \log M + \log(\log(3(L+1))+1/6) + \log 48
\end{equation}
nats/s/Hz with probability at least $1-(5/6)^N$, where $N$ is the number of randomizations used in Algorithm~\ref{alg:0}.
\end{itemize}
\end{thm}
We relegate the proof to Appendix~\ref{proof:prop1}.  From Theorem~\ref{prop:1}(b), we see that the gap between the BF-AF rate and the SDR rate is on the order of $\log M + \log\log L$ in the worst case.  This implies that the BF-AF scheme may not work well in large-scale MIMO relay systems, where there are either many users or many power constraints.  Such a shortcoming motivates us to search for alternative AF schemes.  In the next section, we shall introduce the SBF-AF framework and propose two SBF-AF schemes that provably outperform the BF-AF scheme.  Before we proceed, however, several remarks are in order.

\smallskip
\noindent
{\it Remark 1}: Chang \emph{et al.}~\cite{chang2008approximation} have studied Problem~(BF) with only the total power constraint
and established a bound similar to~\eqref{eq:gap-bd} on the corresponding gap between the BF-AF rate and the SDR rate.  Theorem~\ref{prop:1}(b) generalizes the result in~\cite{chang2008approximation} by allowing both the total power constraint and the per-antenna power constraints to be present in (BF).

\smallskip
\noindent
{\it Remark 2}: Although Theorem~\ref{prop:1}(b) is presented for sum power and per-antenna power
constraints, 
it can be further generalized to cover the case where the constraints in (BF) are replaced~by
$$ {\bm w}^H {\bm Q}_s {\bm w} \le b_s, \quad s=1,\ldots,S $$
for some arbitrary ${\bm Q}_1,\ldots,{\bm Q}_S \in {\mathbb H}_+^{L^2}$ and $b_1,\ldots,b_S\ge0$ (cf.~\eqref{eq:const-cpt} and note from~\eqref{d_v0} and~\eqref{d_v} that ${\bm D}_\ell \in {\mathbb H}_+^{L^2}$ for $\ell=0,1,\ldots,L$).  In particular, it can be shown that the gap between the BF-AF rate and the SDR rate in this case will be on the order of $\log M + \log\log S$.  Such a generalization is useful, as it allows us to model other types of power constraints, such as the interference temperature constraints considered in~\cite{jimulti13}.

\smallskip
\noindent
{\it Remark 3}: It should be noted that in order to practically achieve the BF-AF rate $r_{\sf BF}$, we need to apply a powerful enough channel code with relatively long codelength.

\begin{algorithm}[H]
\caption{Rank-One Gaussian Randomization Procedure for Problem (BF)} \label{alg:0}
\begin{algorithmic}[1]
\STATE input: an optimal solution $\bm{W}^{\star}$ to (SDR), number of randomizations $N \ge 1$
\IF {$\mbox{rank}({\bm W}^\star) \le 1$}
\STATE let ${\bm W}^\star={\bm w}^\star({\bm w}^\star)^H$ and output $\widehat{\bm{w}} = {\bm w}^\star$
\ELSE
\FOR {$n=1$ to $N$}
\STATE generate $\bm{\xi}^n \sim \mathcal{CN}(\bm{0}, \bm{W}^{\star})$
\STATE let 
$$ \widehat{\bm{w}}^n =  \bm{\xi}^n \cdot \min_{\ell=0,1,\ldots,L}\left\{
\sqrt{\frac{\bar{P}_\ell}{{\bm D}_\ell \bullet \left( \bm{\xi}^n \left( {\bm{\xi}^n} \right)^H \right)}} \right\} $$
\STATE set $\theta_n =  \gamma\left( \widehat{\bm{w}}^n(\widehat{\bm{w}}^n)^H \right)$
\ENDFOR
\STATE set $n^\star = \arg\max_{n=1,\ldots,N} \theta_n$ and output $\widehat{\bm{w}} = \widehat{\bm{w}}^{n^\star}$
\ENDIF
\end{algorithmic}
\end{algorithm}

\section{The SBF-AF Schemes} \label{sec:sbf}

\subsection{System Model under the SBF-AF Framework}

The gap between the BF-AF rate and the SDR rate is mainly caused by the fact that the rank-one BF-AF solution $\widehat{\bm W}=\widehat{\bm w}\widehat{\bm w}^H$ does not fully capture the spatial information contained in the potentially high-rank optimal solution ${\bm W}^\star$ to (SDR).  This motivates us to propose the SBF-AF framework to further improve the rate performance. The key idea behind the SBF-AF framework is to adopt time-varying random AF weights, 
{{so that we can simulate ``high-rank'' BF-AF.}}  Specifically, we keep the receive model of the transmitters-to-relay link as in~\eqref{rt}, but modify the AF scheme in~\eqref{eq:xt_v} to
\begin{equation} \label{eq:xt_sbf}
{\bm x}(t) = {\bm V}(t) {\bm r}(t).
\end{equation}
{{Note that unlike the fixed weighting matrix ${\bm V}$ used in the BF-AF scheme~\eqref{eq:xt_v}, the weighting matrix ${\bm V}(t)$ used in~\eqref{eq:xt_sbf} depends on the time $t$.}}

Now, let $\bm{\Omega} \in {\mathbb H}_+^{L^2}$ be a positive semidefinite matrix and $\mathcal{D}=\mathcal{D}(\bm{\Omega})$ be a probability distribution with mean vector $\bm{0}$ and covariance matrix $\bm{\Omega}$.  The choice of $\bm{\Omega}$ and $\mathcal{D}$ will be specified later.  At each time $t$, we generate an independent random vector $\bm{w}(t)$ of AF weights according to the distribution $\mathcal{D}$ and form the AF weighting matrix $\bm{V}(t)$ via $\bm{w}(t)={\rm vec}({\bm V}(t))$.  Since $\bm{w}(t)$ is i.i.d.~in time, we shall drop the time index $t$ and simply write $\bm{w}$ for $\bm{w}(t)$ in the sequel.  Using~\eqref{eq:xt_v} and~\eqref{eq:xt_sbf}, we can rewrite the SISO model in~\eqref{yt} as
\begin{align} 
y_{k,i}(t) &= {\bm g}_{k,i}^H{\bm x}(t) + v_{k,i}(t) \nonumber\\
&= {\bm g}_{k,i}^H{\bm V}(t){\bm f}_ks_k(t) + {\bm g}_{k,i}^H{\bm V}(t)\left(\sum_{m \neq k} {\bm f}_ms_m(t)\right) \nonumber\\
&\quad+ {\bm g}_{k,i}^H{\bm V}(t){\bm n}(t) + {v_{k,i}}(t). \label{sisomodel}
\end{align}
The above expression suggests that we are dealing with a multi-user fast-fading interference channel, where the fading effect is due to the time-varying nature of the AF scheme~\eqref{eq:xt_sbf}.  By treating the interference as noise (cf.~\cite{motahari2009capacity,shang2009new,annapureddy2009gaussian,chaaban2012sub,GNAJ13}), we may define the SBF-AF rate as
\begin{align}
& r_{\sf SBF}(\mathcal{D}) \nonumber \\
&= \min_{k=1,\ldots,G \atop i=1,\ldots,m_k} \mathbb{E}_{\bm{w}\sim\mathcal{D}}\left[\log\left(1+\frac{{\bm w}^H {\bm A}_{k, i}{\bm w}}{{\mathbb{E}_{\bm{w}\sim\mathcal{D}}[{\bm w}^H{\bm C}_{k,i}{\bm w}]}+1}\right)\right]. \label{eq:lowerrate}
\end{align}
In particular, the term ${\mathbb E}_{\bm{w}\sim\mathcal{D}}\left[ {\bm w}^H{\bm C}_{k,i}{\bm w} \right] = {\bm C}_{k,i} \bullet \bm{\Omega}$, which arises from the interference to user $i$ in group $k$, is regarded as the noise variance.

\subsection{The Gaussian and Elliptic SBF-AF Schemes}

With the above setup, it is natural to choose the covariance matrix $\bm{\Omega}$ and probability distribution $\mathcal{D}$ jointly so that the SBF-AF rate defined in~\eqref{eq:lowerrate} is maximized.  However, such a joint optimization problem does not seem to be tractable.  To circumvent this difficulty, one idea is to take a simple zero-mean distribution $\mathcal{D}$ that can be completely characterized by the covariance matrix $\bm{\Omega}$ and then optimize over $\bm{\Omega}$.  Such an idea turns out to be viable and leads to two easily implementable SBF-AF schemes.  The first is the \emph{Gaussian SBF-AF scheme}, where we take $\mathcal{D}$ to be the circularly symmetric complex Gaussian distribution ${\mathcal{CN}}({\bf 0}, {\bm \Omega})$ and generate the AF weight vector $\bm{w}$ via
\begin{equation} \label{eq:w_gau}
{\bm w} \sim {\mathcal{CN}}({\bf 0}, {\bm \Omega}).
\end{equation}
The second is the \emph{elliptic SBF-AF scheme}, where we take $\mathcal{D}$ to be the so-called complex elliptic distribution with mean vector $\bm{0}$ and covariance matrix $\bm{\Omega}$ and generate the AF weight vector $\bm{w}$ via
\begin{equation} \label{eq:w_ellip}
\bm{w} = \frac{ {\bm L}^H \bm{\alpha} }{ \| \bm{\alpha} \| / \sqrt{r} }, \quad \bm{\alpha} \sim \mathcal{CN}( {\bm 0}, {\bm I}_r ),
\end{equation}
where ${\bm L} \in {\mathbb C}^{r\times L}$ satisfies ${\bm L}^H{\bm L} = {\bm \Omega}$ and $r=\mbox{rank}({\bm \Omega})$.  It is known that the random vector $\bm{w}$ in~\eqref{eq:w_ellip} indeed has the prescribed mean vector and covariance matrix; see, e.g.,~\cite{Bok:Multivariate90}.  

To complete the description of the Gaussian and elliptic SBF-AF schemes, it remains to specify the choice of the covariance matrix $\bm{\Omega}$.  Towards that end, consider the following optimization problem, which aims at finding an $\bm{\Omega}$ such that the SBF-AF rate $r_{\sf SBF}$ is maximized, while the power used by the relay antennas, when averaged over all possible realizations of the AF weight vector $\bm{w}$, is below certain prescribed thresholds:
\begin{align*}
({\rm SBF}) \quad \displaystyle \max_{{\bm \Omega} \in \mathbb{H}_+^{L^2}} \quad & r_{\sf SBF}(\mathcal{D}) \\ 
\text{subject to} \,\,\, & \mathbb{E}_{{\bm w}\sim\mathcal{D}}\left[ {\bm w}^H{\bm D}_\ell {\bm w} \right] \le \bar{P}_{\ell}, \,\,\, \ell=0,1,\ldots,L.
\end{align*}
Here, $\bm{D}_0$ and $\bm{D}_\ell$, where $\ell=1,\ldots,L$, are defined in~\eqref{d_v0} and~\eqref{d_v}, respectively; $\mathcal{D}$ is either the circularly symmetric complex Gaussian distribution (which corresponds to the Gaussian SBF-AF scheme) or the complex elliptic distribution (which corresponds to the elliptic SBF-AF scheme) with mean vector $\bm{0}$ and covariance matrix $\bm{\Omega}$.  The upshot of the above formulation is that its optimal solution can be explicitly characterized:
\begin{prop} \label{prop:2}
For both the Gaussian and elliptic SBF-AF schemes, an optimal solution to (SBF) is given by $\bm{W}^\star$, the optimal solution to (SDR).
\end{prop}
The proof of Proposition~\ref{prop:2} can be found in Appendix~\ref{proof:prop2}.  Proposition~\ref{prop:2} shows that by setting $\bm{\Omega}=\bm{W}^\star$, the random AF weight vector $\bm{w}$ satisfies ${\mathbb E}_{\bm{w}\sim\mathcal{D}}\left[ \bm{w}\bm{w}^H \right] = \bm{W}^\star$, which suggests that the proposed SBF-AF schemes are simulating a ``high-rank'' BF-AF scheme.  Moreover, it opens up the possibility of comparing the rates of the proposed SBF-AF schemes with the SDR rate.  In particular, we have the following theorem, which constitutes one of the main results of this paper:
\begin{thm} \label{thm:1}
Let $r_{\sf SBF}({\sf G})$ and $r_{\sf SBF}({\sf E})$ be the Gaussian and elliptic SBF-AF rates, respectively, when $\bm{\Omega}=\bm{W}^\star$.  Then, we have
\begin{align*}
r_{\sf SDR}- r_{\sf SBF}({\sf G}) \le 0.5772
\end{align*}
and
\begin{align*}
r_{\sf SDR}- r_{\sf SBF}({\sf E}) \le \sum_{k=1}^{r-1} \frac{1}{k} - \log(r) < 0.5772,
\end{align*}
where $r={\rm rank}({\bm W}^\star)$.
\end{thm}
We relegate the proof to Appendix \ref{proof:thm1}. Theorem~\ref{thm:1} is significant, as it shows that the Gaussian SBF-AF rate is at most $0.8317$ bits/s/Hz ($0.5772~{\rm nats}/\log2 = 0.8317 ~{\rm bits}$) less than SDR rate $r_{\sf SDR}$, and that the elliptic SBF-AF rate is even better.  Compared with the BF-AF scheme (see Theorem~\ref{prop:1}(b)), we see that the rate performance of the proposed SBF-AF schemes does not degrade with the number of users in the network or the number of power constraints on the relay antennas.  This suggests that the SBF-AF schemes should outperform the SDR-based BF-AF scheme in large-scale systems.

\subsection{Implementation Issues} \label{subsec:impl}
{
To implement the SBF-AF schemes, there are several practical issues that need to be addressed.  First, all nodes in the network (transmitters, receivers, and relay) should be synchronized. This can be realized by virtue of synchronization signals, just as it is usually done in existing relay networks.
Second, to receive the SBF signals, each receiver needs to know the covariance matrix ${\bm \Omega}$. Such information can be transmitted at the beginning of each data frame as part of the preamble. Third, all the relays and receivers should know the instantaneous AF weights. At first sight, it may seem that we need to repeatedly do the signaling for the AF weights. However, this is not necessary. Indeed, we can simply pre-specify a common random seed in the network before transmission. With the aid of the common random seed, the relay and the receivers can locally generate the same SBF-AF weight at each time slot (this is very similar to reproducing the same random realizations in MATLAB by using the same random seed). Therefore, this is no need to inform the receivers the instantaneous SBF-AF weights. Since all transmit signals are synchronized, the receivers can therefore perform simple coherent symbol reception, demodulation, and channel decoding.  In practice, the SBF-AF schemes are just as efficient as the BF-AF schemes with channel coding (see Remark~3 in Section~\ref{subsec:SDR}).
}
The fourth issue concerns the peak-to-average-power ratio (PAPR) at the relay.  Note that the PAPR here is defined over the time-varying AF weights. In this context, although the Gaussian SBF-AF scheme is interesting from a theoretical viewpoint, it may suffer from high instantaneous peak power, as the Gaussian distribution has unbounded support. In practice, we could truncate the Gaussian signal envelope at the relay to limit the peak power. Nevertheless, this may result in performance degradation. By contrast, the elliptic SBF-AF scheme exhibits a good PAPR. Indeed, using the Courant-Fischer min-max theorem, we can prove the following:
\begin{prop} \label{prop:4}
For the elliptic SBF-AF scheme, we will have
\begin{align*}
\bm{w}^H{\bm D}_\ell\bm{w} &\in \left[ r\lambda_{\rm min}^+ \left( {\bm D}_\ell^{1/2}{\bm W}^\star{\bm D}_\ell^{1/2} \right), \right. \\
&\quad\,\,\, \left. r\lambda_{\rm max} \left( {\bm D}_\ell^{1/2}{\bm W}^\star{\bm D}_\ell^{1/2} \right) \right]
\end{align*}
with probability $1$, where $\ell=0,1,\ldots,L$ (recall that ${\bm D}_0$ is defined in~\eqref{d_v0} and ${\bm D}_1,\ldots,{\bm D}_L$ are defined in~\eqref{d_v}).
\end{prop}
Proposition~\ref{prop:4} implies that the instantaneous transmit power of the elliptic SBF-AF scheme is bounded. 
{

To further investigate the issue of PAPR at the relay, we plot the complementary cumulative distribution function (CCDF) in Figure~\ref{fig:papr} to compare the actual PAPR at each relay antenna for the BF-AF and SBF-AF schemes. The CCDF gives the probability that the PAPR of a data block exceeds a given threshold and is one of the most frequently used criteria for measuring PAPR~\cite{han2005overview}. Herein, we adopt the $64$-QAM modulation scheme and test $10000$ data blocks to get the plots. The horizontal and vertical axes represent the threshold $\gamma$ for the PAPR and the probability that the PAPR of a data block exceeds $\gamma$, respectively. The simulation results show that Gaussian SBF-AF has around $5$dB loss while elliptic SBF-AF has only $2$dB loss in CCDF of the PAPR when compared to BF-AF. However, we get a significant rate performance improvement with the SBF-AF schemes.

\begin{figure}[h]
\centering
\includegraphics[width=0.38\textwidth]{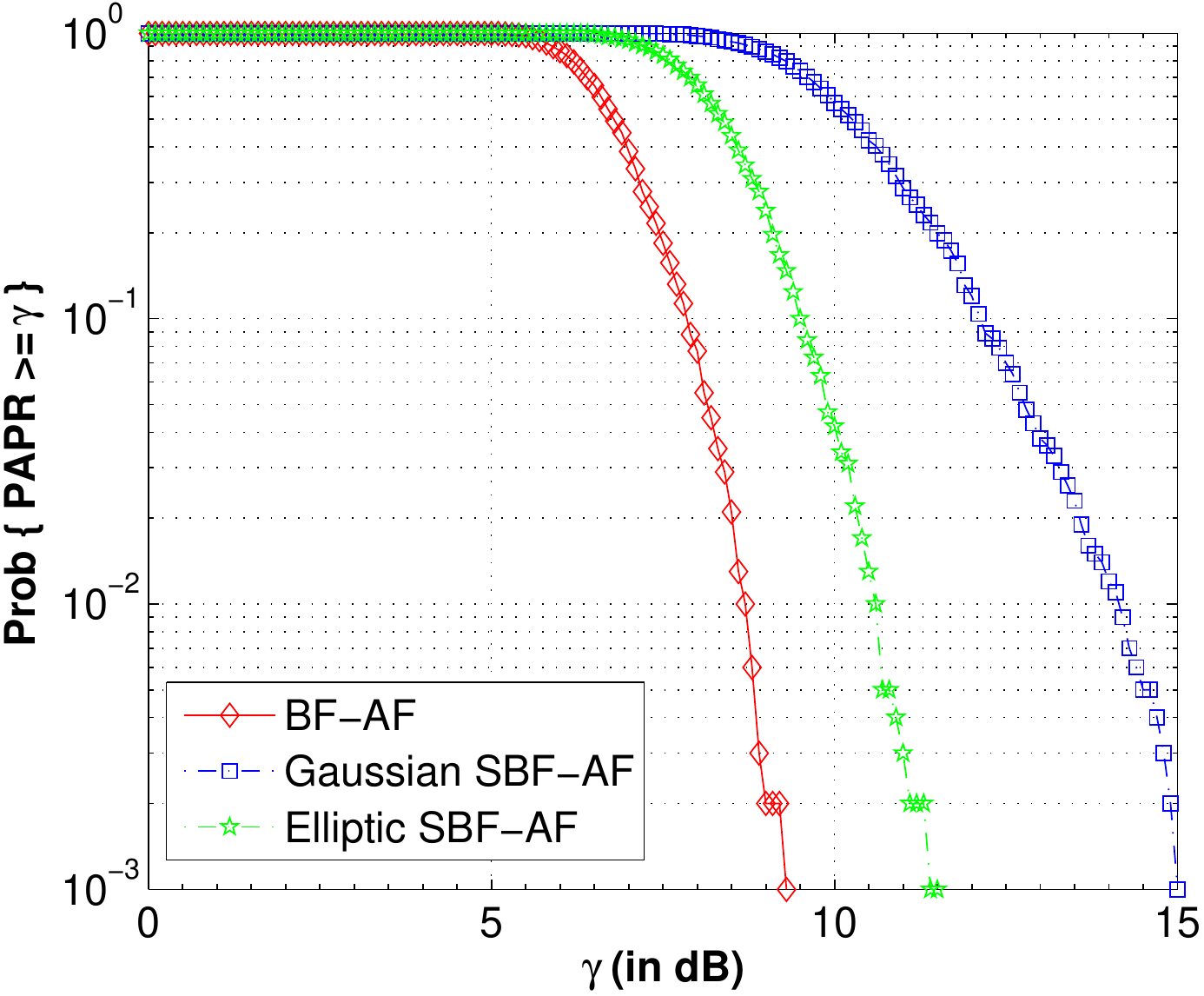}
\caption{The CCDF of the PAPR at each relay for BF-AF and SBF-AF under $64$-QAM modulation.} 
\label{fig:papr}
\end{figure}

}

%
%

\section{Extension to the Distributed Relay Network} \label{sec:DRN}

Although our development so far has focused on the MIMO relay network, it is worth noting that the SBF-AF framework can be applied to other relay networks as well.  As an illustration, let us briefly describe the SBF-AF scheme for a distributed relay network.  The system model of such a network is similar to that of the MIMO relay network described in Section~\ref{subsec:MIMO}, except that the $L$-antenna relay is replaced by $L$ single-antenna relays that are distributively located in the network.  In particular, the received signals cannot be shared among the $L$ relays.  Under this setting, the BF-AF scheme is modeled as
\begin{equation} \label{eq:xt_v_drel}
{\bm x}(t) = {\bm V} {\bm r}(t) \quad\mbox{with}\quad {\bm V} ={\rm Diag}({\bm v}).
\end{equation}
The difference between the BF-AF schemes~\eqref{eq:xt_v} and~\eqref{eq:xt_v_drel} is that the matrix ${\bm V}$ in~\eqref{eq:xt_v_drel} is diagonal, as there is no information exchange among the relays.  Then, similar to the development in Section~\ref{subsec:SDR}, we can formulate the following BF-AF design problem for the distributed relay network:
\begin{equation*}\label{eq:relay_network_MM_re}
\begin{array}{l@{\quad}c@{\quad}l}
({\sf DBF}) & \displaystyle{ \max_{{\bm v} \in \mathbb{C}^L} } & \displaystyle\min_{{k=1,\ldots,G \atop i=1,\ldots,m_k}} \frac{{\bm v}^H \bar{\bm A}_{k,i}{\bm v}}{{\bm v}^H \bar{\bm C}_{k,i}{\bm v}+1} \\
\noalign{\smallskip}
& \text{subject to} & {\bm v}^H {\bm Q}_s{\bm v}\le b_s, \quad s=1,\ldots,S,
\end{array}
\end{equation*}
where
\begin{align*}
\bar {\bm A}_{k,i} &= P_k({\bm f}_k\odot {\bm g}_{k,i}^*)({\bm f}_k\odot {\bm g}_{k,i}^*)^H/\sigma_{k,i}^2, \\ 
\bar {\bm C}_{k,i} &= \sum_{m \neq k} P_m({\bm f}_m\odot {\bm g}_{k,i}^*)({\bm f}_m\odot {\bm g}_{k,i}^*)^H/\sigma_{k,i}^2 \\
&\quad+{\rm Diag} (|g_{k,i}^1|^2\sigma_1^2,\ldots,|g_{k,i}^L|^2\sigma_L^2)/\sigma_{k,i}^2,
\end{align*}
and ${\bm Q}_s$ is the matrix corresponding to the $s$th power constraint (see Remark 2 in Section~\ref{subsec:SDR}). It can be readily seen that Problem (DBF) has exactly the same form as Problem (BF). Hence, the development and analysis of the SDR-based BF-AF scheme and SBF-AF schemes in Sections~\ref{sec:BF} and~\ref{sec:sbf} can be carried over to the distributed relay network directly. We refer the readers to our recent conference paper~\cite{WuLiMaSo2015SPAWC} for details.

\section{Numerical Simulations}\label{sec:sim}
In this section, we provide numerical results to compare the performance of the various AF schemes. Without loss of generality, we assume that each multicast group has an equal number of users (i.e., $m_k=M/G$ for $k=1,\ldots,G$).  The channels ${\bm f}_k, {\bm g}_{k,i}$, where $k=1,\ldots,G$ and $i=1,\ldots,m_k$, are independently generated according to $\mathcal{CN}({\bm 0}, {\bm I})$.  The signal power at each transmitter is $0$dB (i.e., $P_j=0$dB for $j=1,\ldots,G$).  We assume without loss of generality that all antennas of the relay have the same noise power (i.e., $\sigma_{\ell}^2=\sigma_{\sf ant}^2$ for some $\sigma_{\sf ant}^2>0$, where $\ell=1,\ldots,L$), and that all users have the same noise power (i.e., $\sigma_{k,i}^2=\sigma_{\sf user}^2$ for $k=1,\ldots,G$ and $i=1,\ldots,m_k$).  The total power threshold at the relay is $\bar{P}_0$; the power threshold at the $\ell$th antenna of the relay is $\bar{P}_\ell$, where $\ell=1,\ldots,L$. For each AF scheme, $100$ channel realizations were averaged to get the plots. The number of randomizations for generating BF-AF weights is $1000$. {{Note that the channels ${\bm f}_k, {\bm g}_{k,i}$ are fixed for a whole data frame transmission. For the BF-AF scheme~\eqref{eq:xt_v}, a fixed AF weight is adopted; for the SBF-AF scheme~\eqref{eq:xt_sbf}, $T$ time-varying AF weights are generated (here, we assume that the data frame contains $T$ symbols).}} In the following, we will show the numerical results first for the MIMO relay network in Sections~\ref{subsec:tot-pow} to~\ref{subsec:BER} and then for the distributed relay network in Section~\ref{subsec:DRN}.

\subsection{Multicast Rates versus Total Power Threshold at the MIMO Relay} \label{subsec:tot-pow}
In this simulation, we consider the scenario where only the total power constraint is present.  There are $L=8$ antennas at the MIMO relay and $G=2$ multicast groups with a total of $M=16$ users.  In particular, each multicast group has $8$ users.  We set $\sigma_{\sf ant}^2=\sigma_{\sf user}^2=1$ and vary the total power threshold $\bar{P}_0$ at the relay to study the performance of different AF schemes.  The results are shown in Figure \ref{fig:1}.  From the figure, we see that the SDR rate serves as a performance upper bound for the other schemes. The Gaussian SBF-AF scheme outperforms the SDR-based BF-AF scheme when $\bar{P}_0 < 7$dB, while the elliptic SBF-AF scheme outperforms the BF-AF scheme at all the considered power thresholds.  

\begin{figure}[htb]
\centering
\includegraphics[width=0.38\textwidth]{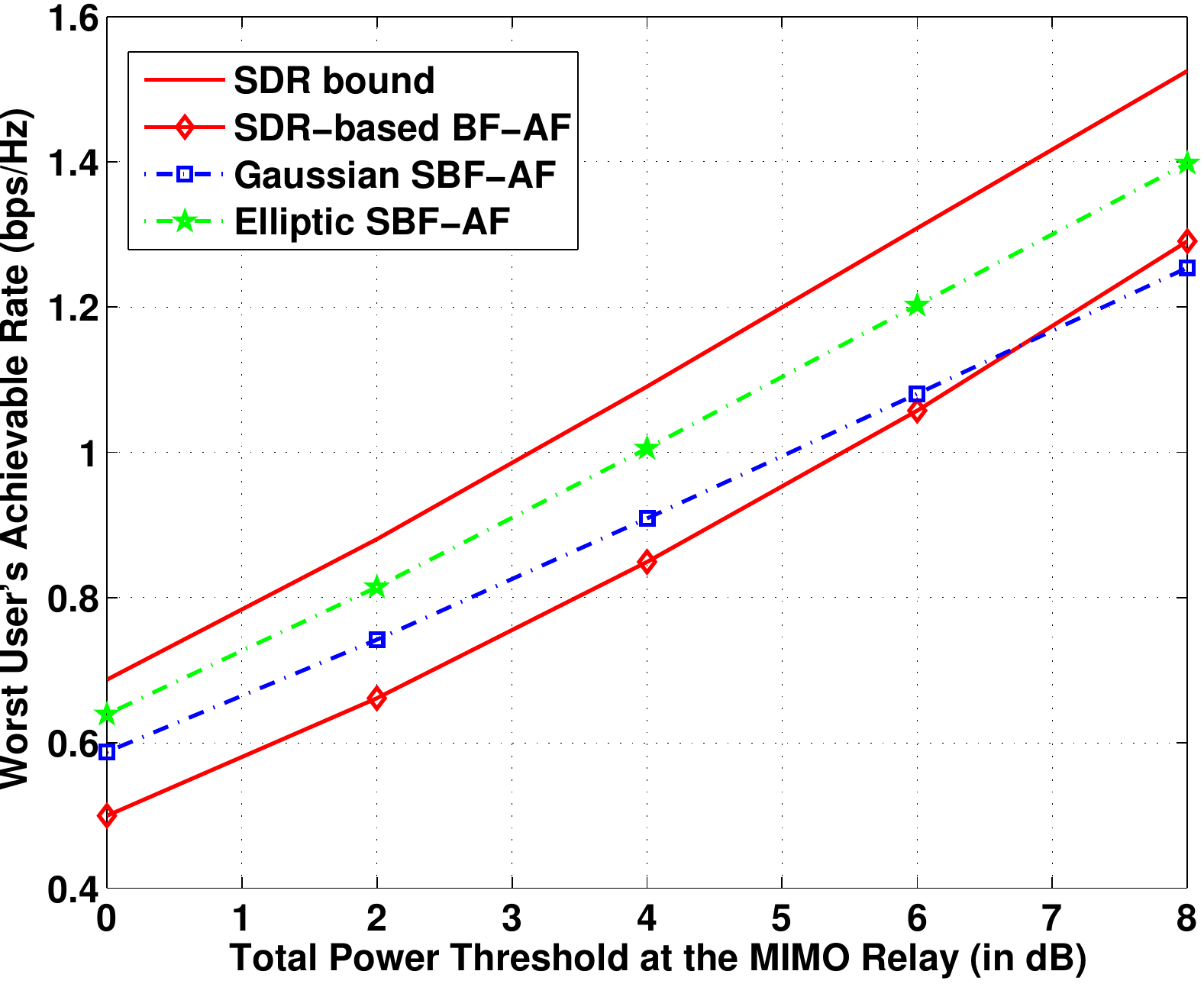}
\caption{Worst user's rate achieved by different AF schemes versus total power threshold at the MIMO relay: $L=8$, $G=2$, $M=16$, $\sigma_{\sf ant}^2=\sigma_{\sf user}^2=1$.}
\label{fig:1}
\end{figure}

\vspace{-\baselineskip}

\subsection{Multicast Rates versus Per-Antenna Power Threshold at the MIMO relay}
In this simulation, we consider the scenario where both total power constraint and per-antenna power constraints are present.  There are $L=4$ antennas at the MIMO relay and $G=1$ multicast group with a total of $M=16$ users.  We set $\sigma_{\sf ant}^2=\sigma_{\sf user}^2=0.25$, and the total power threshold is $\bar{P}_0=3$dB.  We assume that the per-antenna power thresholds are the same for all antennas (i.e., $\bar{P}_1=\cdots=\bar{P}_L$) and vary this threshold to study the performance of different AF schemes.  From Figure~\ref{fig:2}, we see that as the per-antenna power threshold increases, the BF-AF rate and the Gaussian and elliptic SBF-AF rates increase. The SDR rate still serves as a performance upper bound for the other schemes.  On the other hand, the SBF-AF schemes outperform the SDR-based BF-AF scheme at all the considered per-antenna power thresholds.


\begin{figure}
\centering
\includegraphics[width=0.38\textwidth]{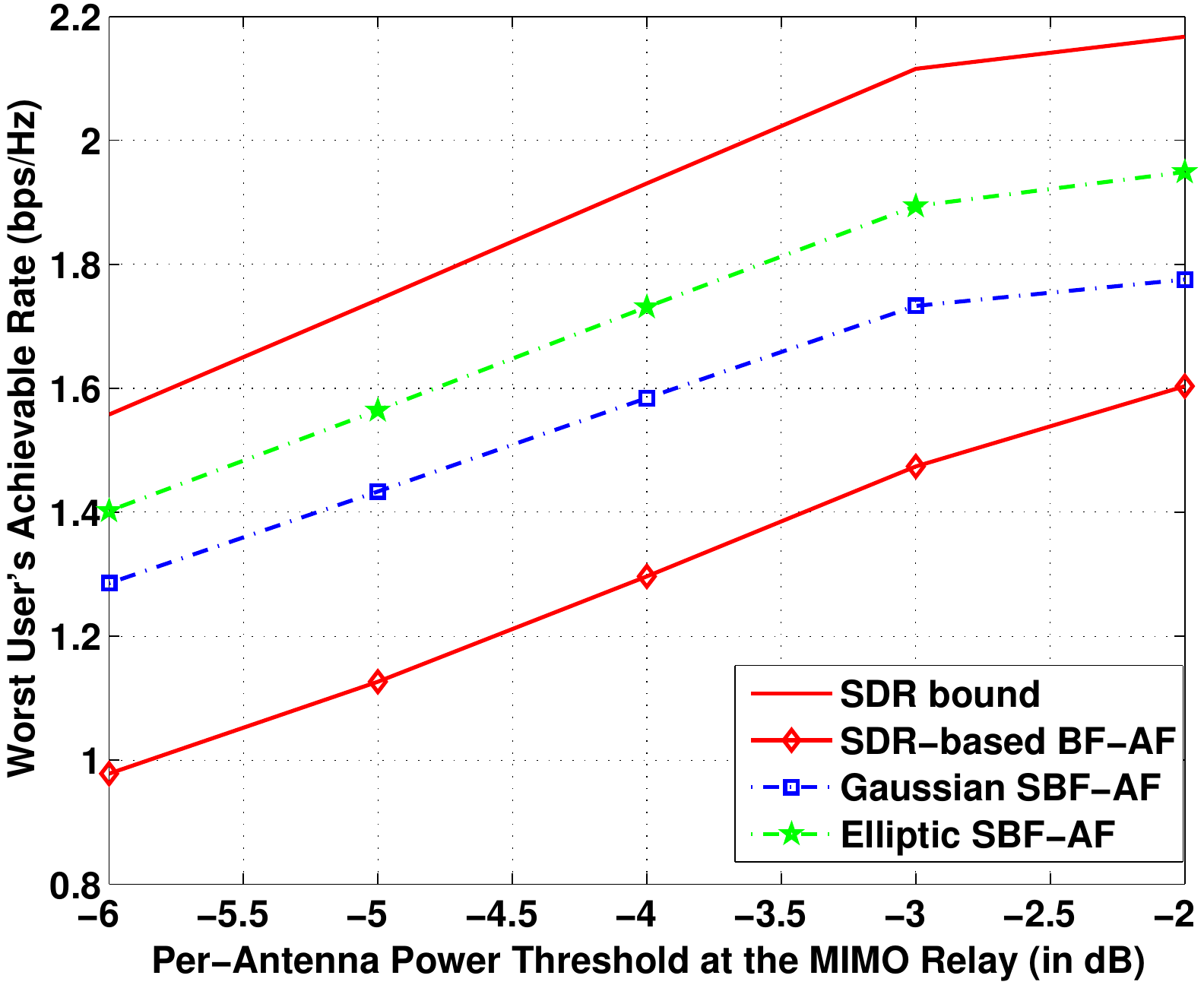}
\caption{Worst user's rate achieved by different AF schemes versus per-antenna power threshold at the MIMO relay: $L=4$, $G=1$, $M=16$, $\bar{P}_0=3$dB, $\sigma_{\sf ant}^2=\sigma_{\sf user}^2=0.25$.} 
\label{fig:2}
\end{figure}

\subsection{Multicast Rates versus Number of Users}
In this simulation, we consider the scenario where only the total power constraint is present.  There are $L=8$ antennas at the MIMO relay and $G=2$ multicast groups.  We set $\sigma_{\sf ant}^2=\sigma_{\sf user}^2=0.25$, and the total power threshold is $\bar{P}_0=6$dB.  In Figure~\ref{fig:3}, we show how the BF-AF rate and the Gaussian and elliptic SBF-AF rates scale with the total number of users $M$.  From the figure, we see that the SDR rate is a performance upper bound for the other schemes.  The BF-AF rate diverges from the SDR rate as $M$ increases.  Moreover, the Gaussian SBF-AF scheme outperforms the SDR-based BF-AF scheme when $M > 10$, while the elliptic SBF-AF scheme outperforms both the SDR-based BF-AF scheme and the Gaussian SBF-AF scheme for all values of $M$. {{Note that when $M$ is small, Problem (SDR) is likely to have a rank-one optimal solution. If it does, then the rank-one solution is also optimal for~(BF). In our experiments, we observe that when $M\le10$, a large number of problem instances do possess a rank-one solution.  This explains why the BF-AF scheme outperforms the Gaussian SBF scheme when $M\le10$.}} It is also worth noting that the Gaussian and elliptic SBF-AF rates exhibit the same scaling as the SDR rate, which is consistent with the results in Theorem~\ref{thm:1}. 

\begin{figure}[htb]
\centering
\includegraphics[width=0.38\textwidth]{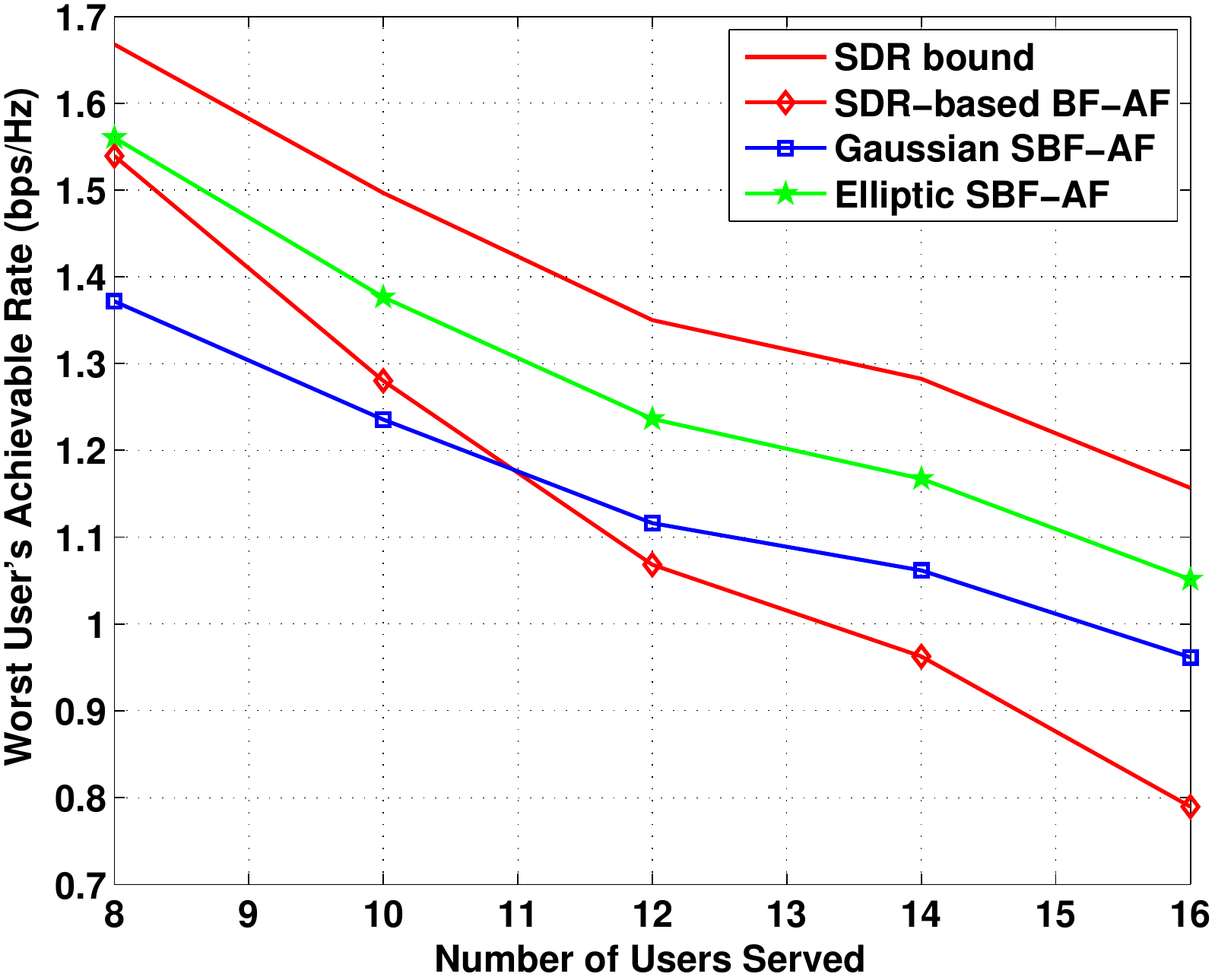}
\caption{Worst user's rate achieved by different AF schemes versus number of users in the MIMO relay system: $L=8$, $G=2$, $\bar{P}_0=6$dB, $\sigma_{\sf ant}^2=\sigma_{\sf user}^2=0.25$.}
\label{fig:3}
\end{figure}

\vspace{-\baselineskip}

\subsection{Multicast Rates versus Number of Power Constraints}
In this simulation, we consider the scenario where both total power constraint and per-antenna power constraints are present.  There are $L=4$ antennas at the MIMO relay and $G=1$ multicast group with a total of $M=16$ users.  We set $\sigma_{\sf ant}^2=\sigma_{\sf user}^2=0.25$, and the total power threshold is $\bar{P}_0=4$dB.  We assume that the per-antenna power threshold is $-5$dB for all antennas (i.e., $\bar{P}_1=\cdots=\bar{P}_L=-5$dB) and vary the number of per-antenna power constraints from $0$ to $L$ to study the performance of different AF schemes.  Figure~\ref{fig:4} shows that the BF-AF rate and the Gaussian and elliptic SBF-AF rates are still upper bounded by the SDR rate.  As the number of per-antenna power constraints increases, the BF-AF rate diverges from the SDR rate, while the Gaussian and elliptic SBF-AF rates exhibit the same scaling as the SDR rate.  Moreover, the Gaussian SBF-AF scheme outperforms the SDR-based BF-AF scheme when the number of per-antenna power constraints is greater than $2$, while the elliptic SBF-AF scheme always outperforms the BF-AF scheme, regardless of the number of per-antenna power constraints.

\begin{figure}[htb]
\centering
\includegraphics[width=0.38\textwidth]{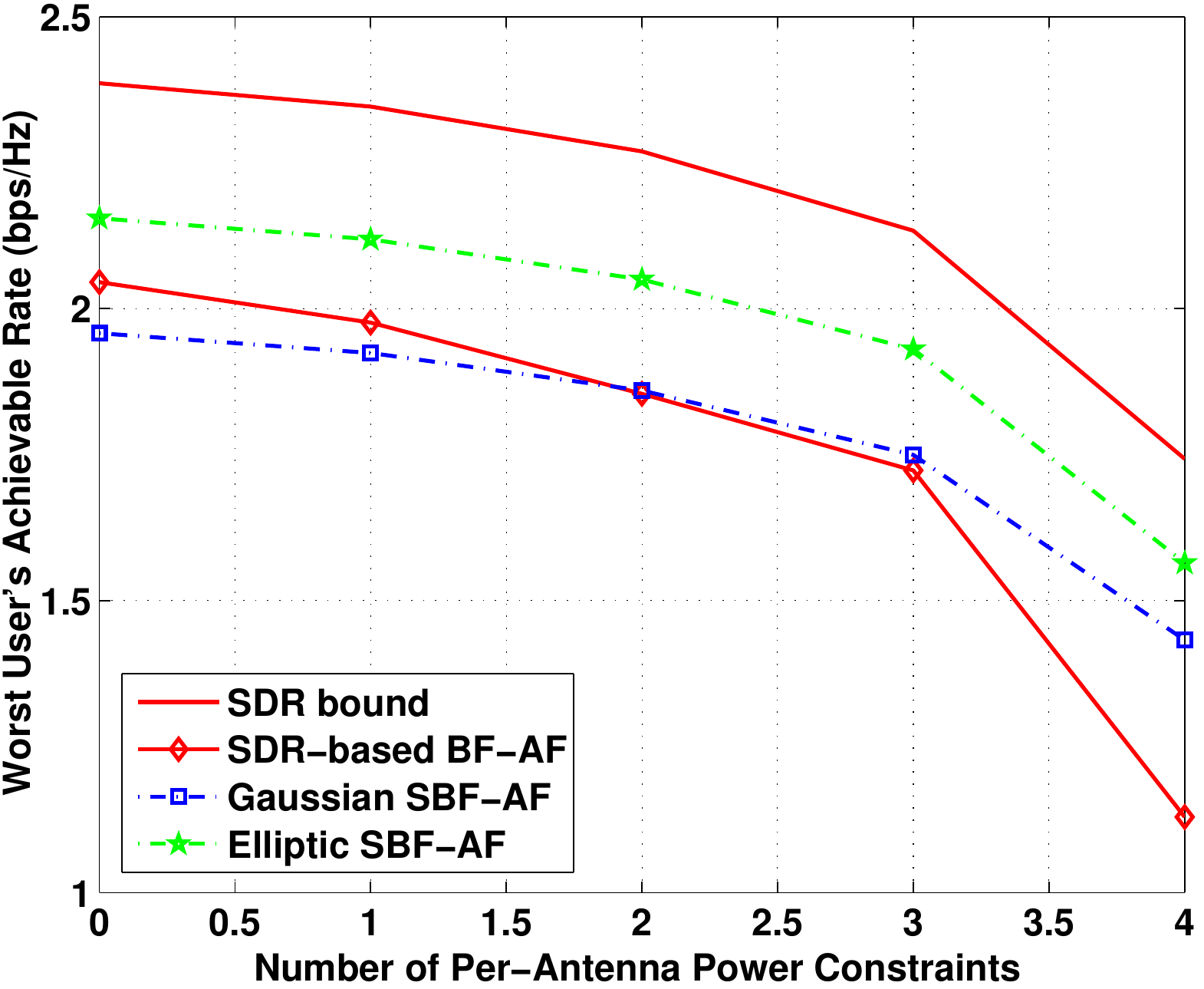}
\caption{Worst user's rate achieved by different AF schemes versus number of per-antenna power constraints: $L=4$, $G=1$, $M=16$, $\bar{P}_0=4$dB, $\bar{P}_\ell = -5$dB for $\ell=1,\ldots,L$, $\sigma_{\sf ant}^2=\sigma_{\sf user}^2=0.25$.}
\label{fig:4}
\end{figure}

\vspace{-\baselineskip}

\subsection{Actual Bit Error Rate (BER) Performance} \label{subsec:BER}
To further demonstrate the efficacy of the proposed SBF-AF schemes, we consider again the scenario in Section~\ref{subsec:tot-pow} and study the coded bit error rate (BER) performance of the different AF schemes.  {The system setting here is $L=8$, $G=2$, $M=16$, and $\sigma_{\sf ant}^2=\sigma_{\sf user}^2=1$, just like that in Figure \ref{fig:1}. For each symbol time slot, we simulate the actual AF process by generating $s_j(t), {n}^\ell(t)$ according to the receive models~\eqref{yt} and \eqref{sisomodel}. In particular,  the SBF weighting matrix ${\bm V}(t)$ in \eqref{sisomodel} is generated for each of the symbol time slot $t$ following \eqref{eq:w_gau} or \eqref{eq:w_ellip}. We then perform coherent detection and iterative decoding on $s_j(t)$ at each receiver. }  The resulting BERs are shown in Figures~\ref{fig:5} and~\ref{fig:6}.  To simulate the SDR bound in the BER plots, we assume that there exists an SISO channel whose SINR is equal to $\gamma({\bm W}^\star)$. In our simulations, we adopt a gray-coded QPSK modulation scheme and a rate-$1/3$ turbo code in \cite{Std:16e} with codelengths $2880$ and $576$.  We simulate $100$ code blocks for each channel realization and thus the BER reliability level is $10$e$-4$. From Figure~\ref{fig:5}, we see that under a relatively long codelength, the actual BER performance of the SBF-AF schemes outperform the SDR-based BF-AF scheme at almost all power thresholds.  Moreover, the elliptic SBF-AF scheme achieves the best BER performance, which is consistent with the results in Figure~\ref{fig:1}. When the channel codelength is relatively short, Figure~\ref{fig:6} shows that the BER performance of the Gaussian SBF-AF scheme degrades a bit, while the elliptic SBF-AF scheme can still outperform the SDR-based BF-AF scheme.  The results in Figures~\ref{fig:1}, \ref{fig:5} and~\ref{fig:6} imply that the SBF-AF schemes, especially the elliptic SBF-AF scheme, can achieve a good rate and are more effective than the existing SDR-based BF-AF scheme. The advantage of the SBF-AF schemes becomes even more apparent when there are many users in the MIMO relay system.

\begin{figure}[htb]
\centering
\includegraphics[width=0.38\textwidth]{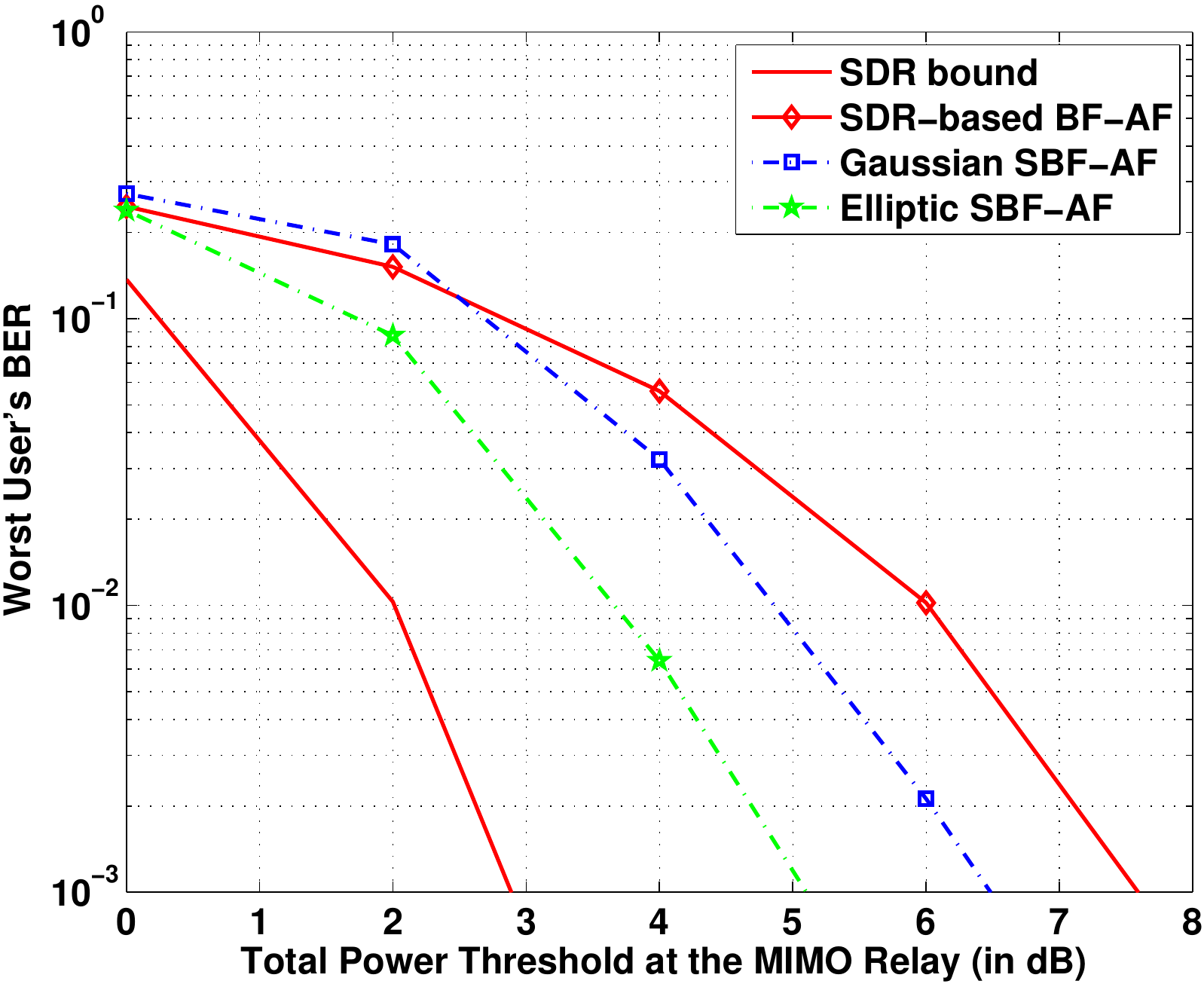}
\caption{Worst user's BER achieved by different AF schemes versus total power threshold at the MIMO relay: $L=8$, $G=2$, $M=16$, $\sigma_{\sf ant}^2=\sigma_{\sf user}^2=1$. A rate-$\frac{1}{3}$ turbo code with codelength $2880$ is used.}
\label{fig:5}
\end{figure}

\begin{figure}[htb]
\centering
\includegraphics[width=0.38\textwidth]{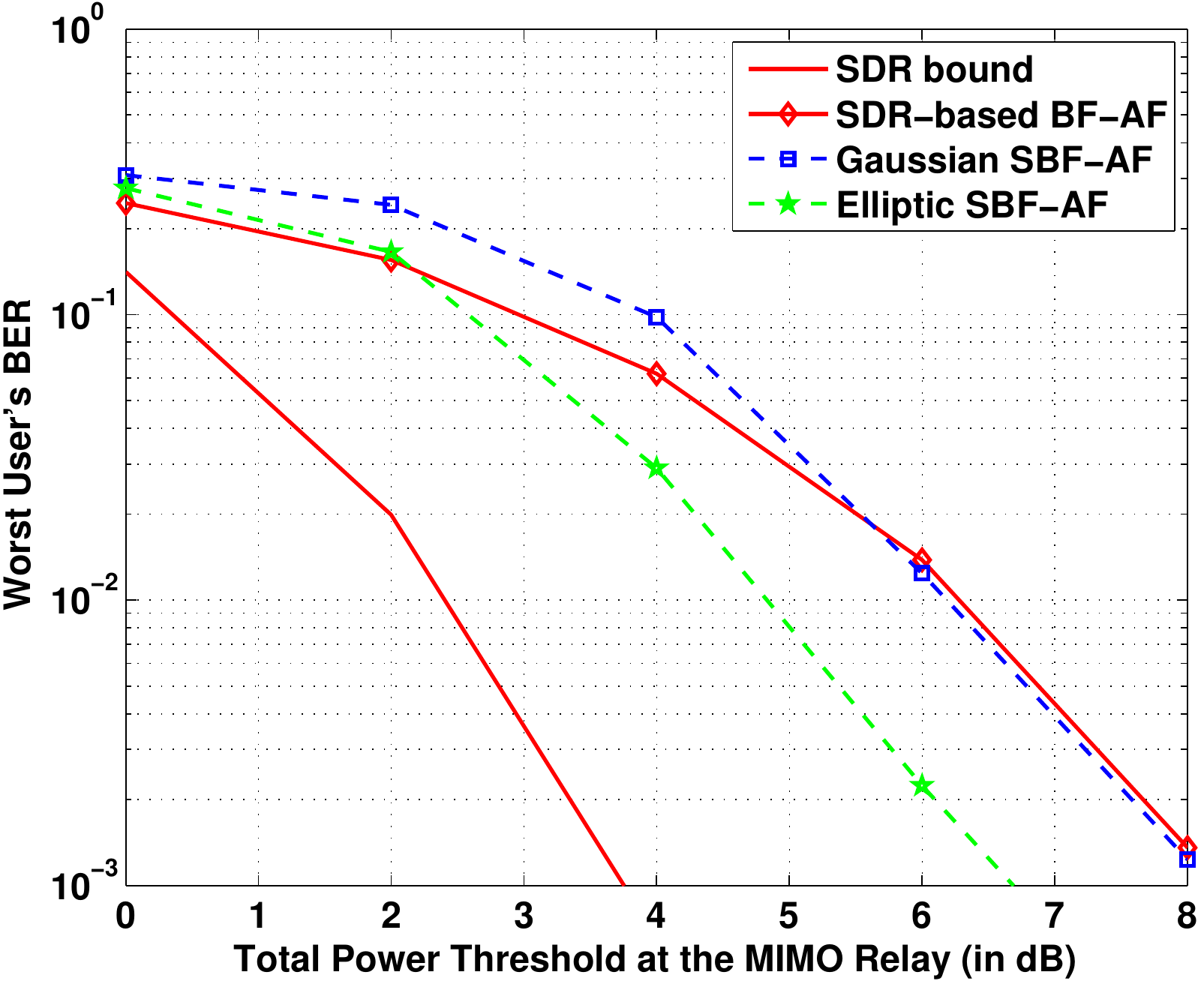}
\caption{Worst user's BER achieved by different AF schemes versus total power threshold at the MIMO relay: $L=8$, $G=2$, $M=16$, $\sigma_{\sf ant}^2=\sigma_{\sf user}^2=1$. A rate-$\frac{1}{3}$ turbo code with codelength $576$ is used.}
\label{fig:6}
\end{figure}

\subsection{Simulation Results for a Distributed Relay Network} \label{subsec:DRN}
In this section, we provide numerical results to demonstrate the effectiveness of our proposed SBF-AF schemes in a distributed relay network. The setting is essentially the same as that in the MIMO relay network, except that the multiple single-antenna relays do not share the received signals. For simplicity, we consider the scenario where only the total power constraint is present in Problem (DBF) (i.e., $S=1$ and ${\bm Q}_1={\bm I}$).  There are $L=8$ relays and $G=2$ multicast groups in the distributed relay network.  We set $\sigma_{\sf ant}^2=\sigma_{\sf user}^2=0.25$.  Figure~\ref{fig1} shows how the BF-AF rate and the Gaussian and elliptic SBF-AF rates scale with the total number of users $M$ when the total power threshold is fixed at $6$dB (i.e., $b_1=6$dB in Problem (DBF)).  From the figure, we see that the BF-AF rate diverges from the SDR rate as $M$ increases.  The Gaussian SBF-AF scheme outperforms the SDR-based BF-AF scheme when $M > 10$, while the elliptic SBF-AF scheme outperforms the SDR-based BF-AF scheme for all values of $M$. Moreover, the Gaussian and elliptic SBF-AF rates exhibit the same scaling as the SDR rate, which is consistent with the results obtained for the MIMO relay network. 

\begin{figure}[htb]
\centering
\includegraphics[width = 0.38\textwidth]{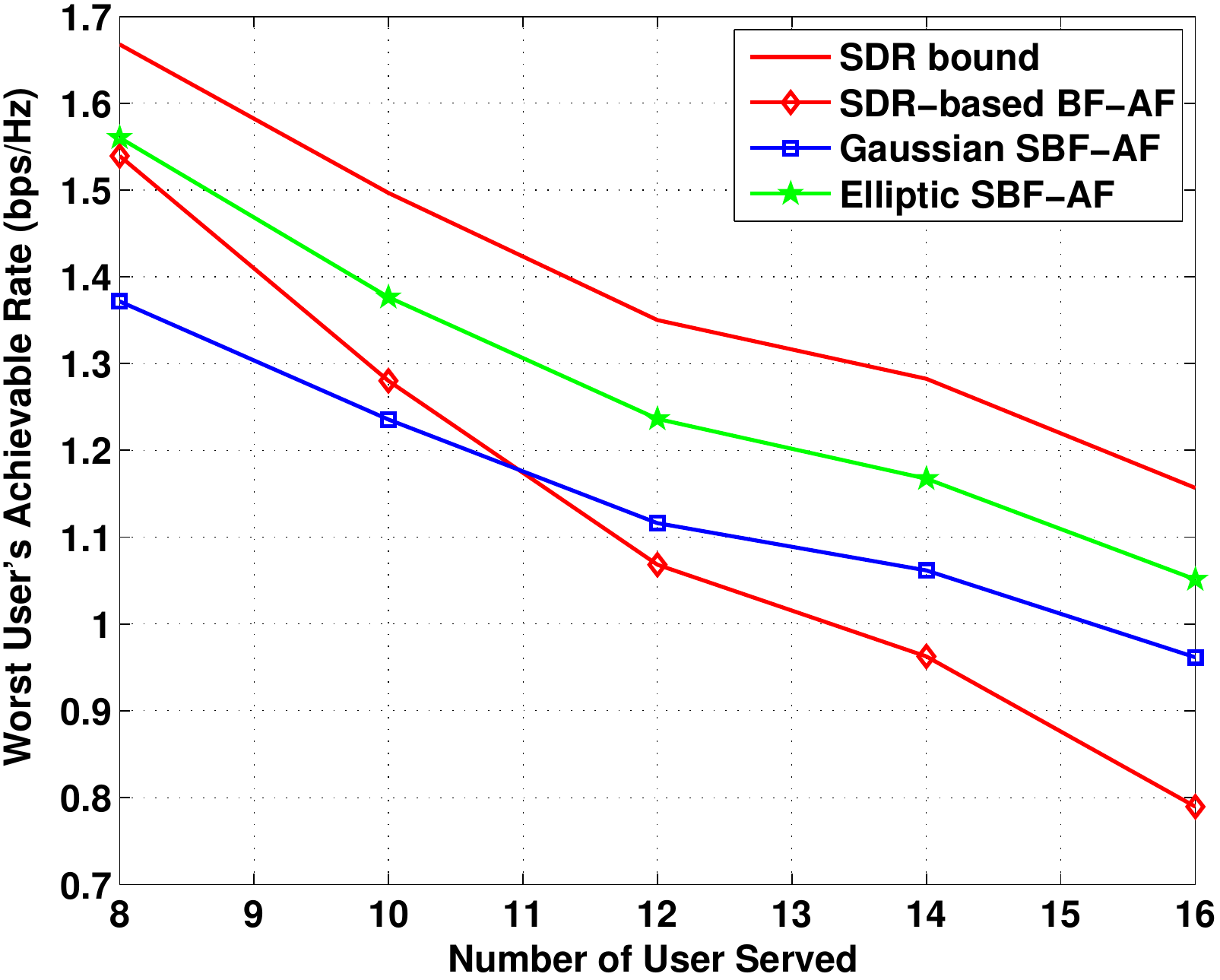}
\caption{Worst user's rate achieved by different AF schemes versus number of users in a distributed relay network.} 
\label{fig1}
\end{figure}

In Figure \ref{fig2}, we compare the coded BER performance of the different AF schemes for the case where $M=12$. Here, we also adopt a gray-coded QPSK modulation scheme and a rate-$1/3$ turbo code in \cite{Std:16e} with codelengths $576$ and $2880$. From the figure, we see that the actual BER performance of the SBF-AF schemes outperform the SDR-based BF-AF scheme at almost all power thresholds, and the elliptic SBF-AF scheme achieves the best BER performance.  The results are consistent with those in Figure~\ref{fig1} and show that the SBF-AF schemes can also achieve a good rate in a distributed relay network.

\begin{figure}[htb]
\centering
\includegraphics[width = 0.38\textwidth]{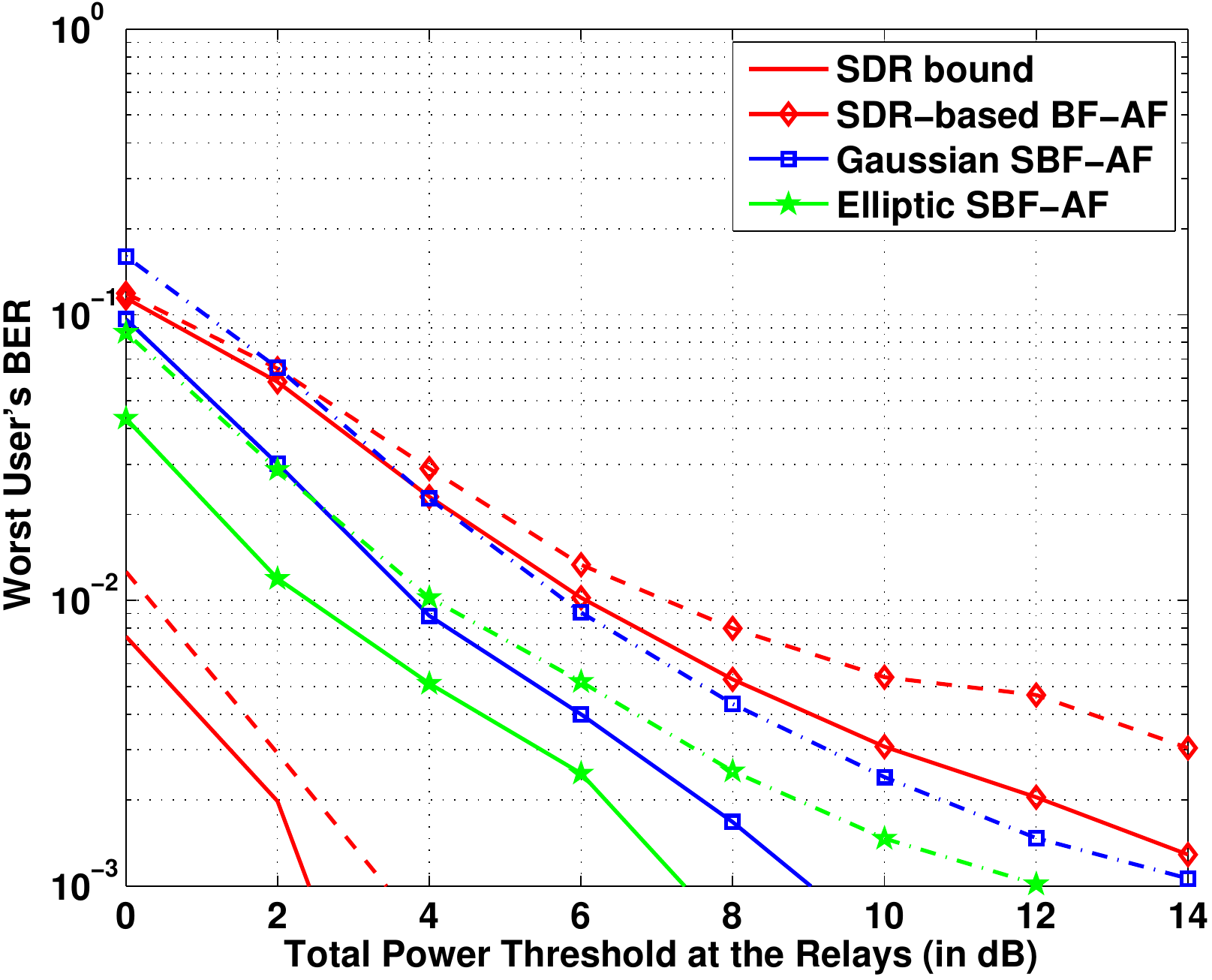}
\caption{Worst user's BER achieved by different AF schemes versus total power threshold in a distributed relay network. The dashed and solid curves correspond to the rate-$\frac{1}{3}$ turbo code with codelengths $576$ and $2880$, respectively.}
\label{fig2}
\end{figure}

{
\subsection{A Comparison with the Feasible Point Pursuit (FPP) Algorithm} \label{subsec:FPP}
In this section, we compare the proposed SBF-AF schemes with the FPP algorithm \cite{mehanna2015feasible,christopoulos2015multicast}, which is recently proposed for solving QCQPs and has been numerically proven to outperform most of the existing algorithms. Specifically, we compare the SBF-AF schemes with the FPP scheme in \cite{mehanna2015feasible} in a distributed relay network and with the FPP-SCA scheme in \cite{christopoulos2015multicast} in an MIMO relay network. In the left sub-figure of Figure \ref{fig:fpp}, we consider only the total power constraint and use the system setting $L=8$, $G=1$, $M=16$, $\sigma_{\sf ant}^2=\sigma_{\sf user}^2=0.25$. In the right sub-figure of Figure \ref{fig:fpp}, we include both the total power constraint and per-antenna power constraints.  The system setting is $L=4$, $G=1$, $M=16$, $\sigma_{\sf ant}^2=\sigma_{\sf user}^2=0.25$, and $\bar{P}_0=3$dB.  We assume that the per-antenna power thresholds are the same for all antennas (i.e., $\bar{P}_1=\cdots=\bar{P}_L$). The results show that the elliptic SBF-AF scheme exhibits a performance gain over the FPP scheme, and both SBF-AF schemes outperform the FPP-SCA scheme. 

\begin{figure}[htb]
\centering
\includegraphics[width = 0.38\textwidth]{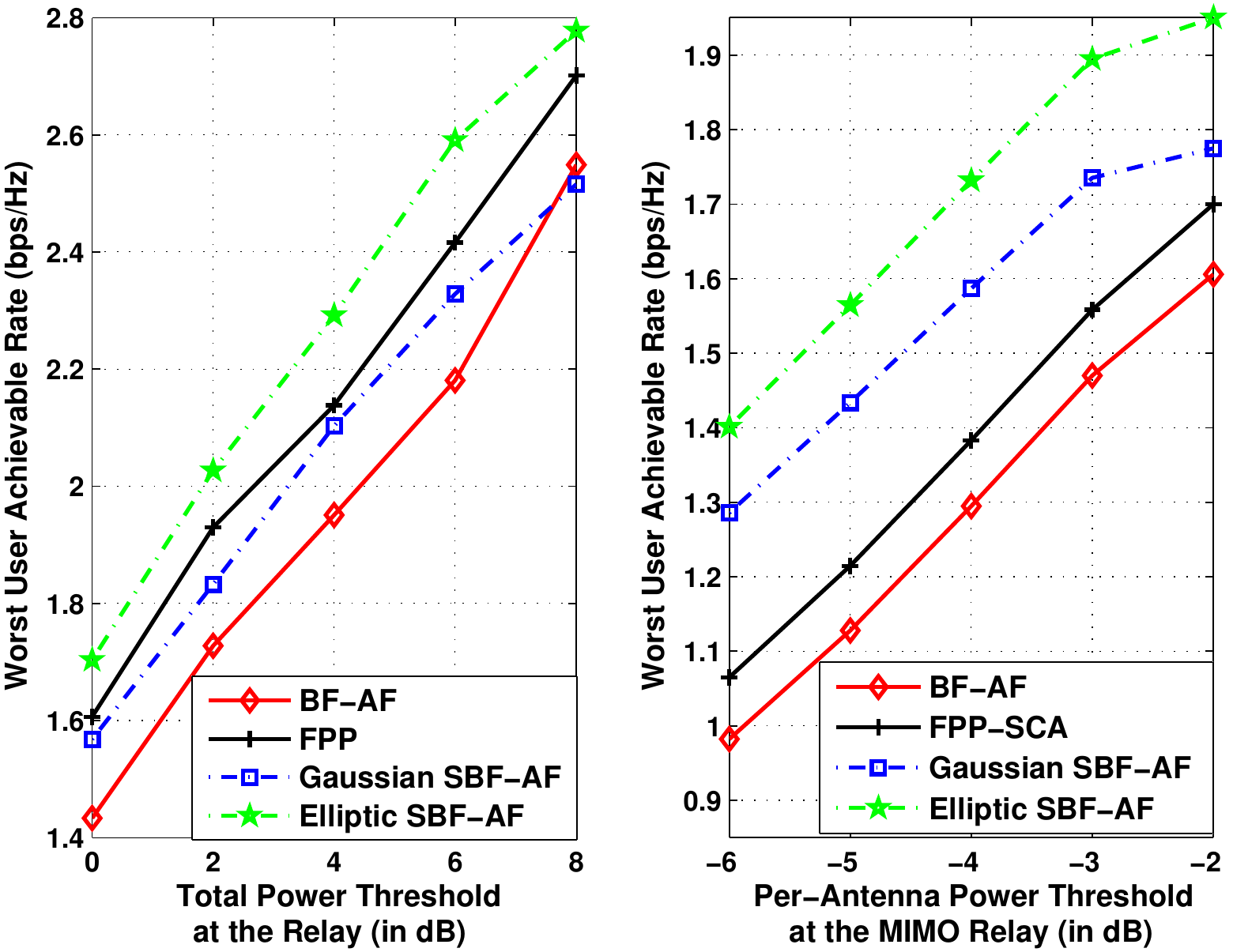}
\caption{Comparison with the feasible point pursuit method.}
\label{fig:fpp}
\end{figure}

}

\section{Conclusions} \label{sec:conclusions}
In this paper, we studied various AF schemes for an MIMO relay network.  We proved that for the classic SDR-based BF-AF scheme, the gap between the BF-AF rate and the SDR rate will grow with the number of users and power constraints.  Thus, the SDR-based BF-AF scheme may not work well for large-scale systems.  In view of this, we proposed two SBF-AF schemes, namely the Gaussian and elliptic SBF-AF, to improve the rate performance.  The proposed SBF-AF schemes employ time-varying AF weights and are essentially simulating a ``high-rank'' BF-AF scheme.  As such, they are capable of outperforming the SDR-based BF-AF scheme.  Indeed, we proved that the Gaussian and elliptic SBF-AF rates are at most $0.8317$ bits/s/Hz less than the SDR rate, irrespective of the number of users or power constraints.  We further demonstrated the superiority of the proposed SBF-AF schemes by comparing their BER performance with that of the SDR-based BF-AF scheme.  Lastly, we discussed how the SBF-AF framework can be applied to a distributed relay network and showed the good rate performance of the corresponding SBF-AF schemes.  As the SBF framework proves to be quite powerful, a possible future direction would be to develop SBF-AF schemes for more involved relay networks, such as a two-way relay network with direct link.  It would also be interesting to consider imperfect CSIs in the SBF framework.

\section*{Acknowledgment}
The authors would like to sincerely thank the Editor and the anonymous reviewers
for their helpful and insightful comments, which help improve the quality of the paper. 
Moreover, we want to take this opportunity to express our gratitude to Professor Nikos Sidiropoulos and his group at University of Minnesota for kindly providing their MATLAB code and data to help us produce part of Figure 12.

\appendix

\subsection{Proof of Theorem~\ref{prop:1}} \label{proof:prop1}
Let $\bar{\bm{W}}^\star$ be an optimal solution to (SDR).  Set
\begin{align*}
\gamma_{k,i}^\star &= \frac{\bm{A}_{k,i}\bullet \bar{\bm{W}}^\star}{\bm{C}_{k,i} \bullet \bar{\bm{W}}^\star + 1}, && k=1,\ldots,G, i=1,\ldots,m_k, \\
P_\ell^\star &= \bm{D}_\ell \bullet \bar{\bm{W}}^\star, && \ell=0,1,\ldots,L.
\end{align*}
It is clear that $\bar{\bm{W}}^\star$ is feasible for the following SDP:
\begin{align}
\displaystyle \max_{\bm{W} \in \mathbb{H}_+^{L^2}} \quad & \left( \bm{A}_{1,1} - \gamma_{1,1}^\star \bm{C}_{1,1} \right) \bullet \bm{W} \nonumber \\
\text{subject to} \,\,\, & \left( \bm{A}_{k,i} - \gamma_{k,i}^\star \bm{C}_{k,i} \right) \bullet \bm{W} = \gamma_{k,i}^\star, \,\,\, (k,i)\not=(1,1), \nonumber \\
& \bm{D}_\ell \bullet \bm{W} = P_\ell^\star, \,\,\, \ell=0,1,\ldots,L. \label{eq:aux-sdr}
\end{align}
Moreover, since $\bm{D}_0$ is positive definite, the feasible set of Problem~\eqref{eq:aux-sdr} is compact.  This implies that Problem~\eqref{eq:aux-sdr} has an optimal solution.  Hence, by~\cite[Theorem 5.1]{Jnl:Yongwei_rank}, there exists a rank-one optimal solution $\bm{W}^\star$ to Problem~\eqref{eq:aux-sdr} whenever $M-1+L+1=M+L\le 3$.  Upon observing that $\bm{W}^\star$ is also optimal for (SDR), we obtain the conclusion in Theorem~\ref{prop:1}(a).

To prove Theorem~\ref{prop:1}(b), fix a particular randomization $n\in\{1,\ldots,N\}$ in Algorithm~\ref{alg:0} and let $\widehat{\bm W} = \bm{\xi}^n \left( {\bm{\xi}^n} \right)^H$, where $\bm{\xi}^n \sim \mathcal{CN}(\bm{0},\bm{W}^\star)$.  For any $\beta>0$ and $\rho>1$, consider the events
\begin{align*}
\mathcal{E}_{k,i} &= \left\{ \frac{{\bm A}_{k,i} \bullet {\widehat {\bm W}}}{{\bm C}_{k,i}  \bullet \widehat{\bm{W}} + 1} \le \beta \frac{{\bm A}_{k,i} \bullet {\bm W}^\star}{{\bm C}_{k,i}  \bullet {\bm{W}}^\star + 1} \right\}, \\
\mathcal{F}_\ell &= \left\{ {\bm D}_\ell \bullet \widehat{\bm{W}} \ge \rho {\bm D}_\ell \bullet {\bm{W}}^\star \right\},
\end{align*}
where $k=1,\ldots,G$, $i=1,\ldots,m_k$, and $\ell=0,1,\ldots,L$.  To bound $\Pr(\mathcal{E}_{k,i})$ and $\Pr(\mathcal{F}_\ell)$, we need the following results:
\begin{lemma} \label{lem:3} 
Let ${\bm A}, {\bm C} \in {\mathbb H}_+^{L^2}$ be such that ${\rm rank}({\bm A})=1$.  Then,
$$ \Pr\left( \frac{{\bm A} \bullet \widehat{\bm W}}{{\bm C} \bullet \widehat{\bm W}+1} \le \beta \frac{{\bm A} \bullet {\bm W}^\star}{{\bm C} \bullet {\bm W}^\star+1} \right) \le \frac{3\beta}{1-2\beta}, $$
where $0<\beta<1/2$.
\end{lemma}
\begin{lemma} \label{lem:4} 
Let ${\bm D} \in {\mathbb H}_+^{L^2}$ be given.  If $\bm{D} \bullet \bm{W}^\star=0$, then $\bm{D}\bullet\widehat{\bm{W}} = 0$ almost surely.  Otherwise, for any $\rho>1$, 
\begin{eqnarray*}
\Pr \left( {\bm D} \bullet \widehat{\bm W} \ge \rho {\bm D} \bullet {\bm W}^\star \right) \le \exp\left( -\frac{\rho-1}{6} \right).
\end{eqnarray*}
\end{lemma}
Lemma \ref{lem:3} is a simple consequence of~\cite[Lemma 2]{chang2008approximation}; cf.~\cite[Lemma 2]{jimulti13}.  On the other hand, Lemma~\ref{lem:4} can be obtained from the proof of~\cite[Proposition 2.1]{Jnl:SoYe2008} and the remarks after it.
 
From~\eqref{ak_v}, we have $\mbox{rank}(\bm{A}_{k,i})=1$ for $k=1,\ldots,G$ and $i=1,\ldots,m_k$.  Hence, by taking $\beta=1/(8M)$ and invoking Lemma~\ref{lem:3}, we have $\Pr(\mathcal{E}_{k,i}) \le 3/2(4M-1)$ for $k=1,\ldots,G$ and $i=1,\ldots,m_k$.  This, together with the union bound, yields
$$ \Pr\left( \bigcup_{k=1,\ldots,G \atop i=1,\ldots,m_k} \mathcal{E}_{k,i} \right) \le \sum_{k=1,\ldots,G \atop i=1,\ldots,m_k} \Pr(\mathcal{E}_{k,i}) \le \frac{3M}{2(4M-1)} < \frac{1}{2}. $$
In addition, by taking $\rho=6\log(3(L+1))+1$ and invoking Lemma~\ref{lem:4}, we have
$$ \Pr\left( \bigcup_{\ell=0}^{L} \mathcal{F}_\ell \right) \le \sum_{\ell=0}^L \Pr(\mathcal{F}_\ell) \le (L+1)\cdot\exp\left(-\frac{\rho-1}{6}\right) = \frac{1}{3}. $$
Thus, if we let $\mathcal{E}_{k,i}^c$ (resp.~$\mathcal{F}_\ell^c$) to be the complement of $\mathcal{E}_{k,i}$ (resp.~$\mathcal{F}_\ell$), then
\begin{align*}
& \Pr\left( \left( \bigcap_{k=1,\ldots,G \atop i=1,\ldots,m_k} \mathcal{E}_{k,i}^c \right) \cap \left( \bigcap_{\ell=0}^L \mathcal{F}_\ell^c \right) \right) \\
&\ge 1 - \Pr\left( \bigcup_{k=1,\ldots,G \atop i=1,\ldots,m_k} \mathcal{E}_{k,i} \right) - \Pr\left( \bigcup_{\ell=0}^{L} \mathcal{F}_\ell \right) \\
&\ge \frac{1}{6}.
\end{align*}
In particular, with probability at least $1/6$, the rank-one solution $\widehat{\bm{W}}/\rho$ is feasible for Problem (SDR) and
\begin{align*}
\gamma\left( \widehat{\bm{W}}/\rho \right) &= \min_{k=1,\ldots,G \atop i=1,\ldots,m_k }\frac{{\bm A}_{k,i}\bullet (\widehat{\bm W}/\rho)} {{\bm C}_{k,i}\bullet (\widehat{\bm W}/\rho)+1} \\
&= \min_{k=1,\ldots,G \atop i=1,\ldots,m_k }\frac{{\bm A}_{k,i}\bullet \widehat {\bm W}} {{\bm C}_{k,i}\bullet \widehat{\bm W}+1} \cdot \frac{{\bm C}_{k,i} \bullet \widehat{\bm W}+1} {{\bm C}_{k,i} \bullet \widehat{\bm W}+\rho} \\
&\ge \frac{1}{\rho} \min_{k=1,\ldots,G \atop i=1,\ldots,m_k }\frac{{\bm A}_{k,i} \bullet \widehat {\bm W}} {{\bm C}_{k, i}\bullet \widehat{\bm W}+1} \\
&\ge \frac{\beta}{\rho}  \cdot \gamma\left( {\bm{W}^\star} \right) \\
&= \frac{1}{8M(6\log(3(L+1))+1)}\cdot \gamma\left( {\bm{W}^\star} \right).
\end{align*}
Since this holds for each randomization $n\in\{1,\ldots,N\}$, it follows that
\begin{align*}
& \Pr \left( \left\{ \exists n: \gamma\left( \widehat{\bm{w}}^n(\widehat{\bm{w}}^n)^H \right) \ge \frac{\gamma\left( {\bm{W}^\star} \right)}{8M(6\log(3(L+1))+1)} \right\} \right) \\
& \ge 1-(5/6)^N.
\end{align*}
Using the above result and the monotonicity of the logarithm, we see that with probability at least $1-(5/6)^N$,
\begin{align*}
& r_{\sf SDR} - r_{\sf BF}  \\
&= \log\left( 1+\gamma(\bm{W}^\star) \right) - \max_{n=1,\ldots,N} \log\left( 1+\gamma\left( \widehat{\bm w}^n (\widehat{\bm w}^n)^H \right) \right) \\
&\le \log\left(\frac{1+\gamma( {\bm W}^\star )}{1+(\beta/\rho)\gamma\left( {\bm W}^\star \right)}\right) \\
&\le \log(8M(6\log(3(L+1))+1)) \\
&= \log M + \log(\log(3(L+1))+1/6) + \log 48.
\end{align*}
This completes the proof of Theorem~\ref{prop:1}(b).
\hfill $\blacksquare$

\subsection{Proof of Proposition~\ref{prop:2}} \label{proof:prop2}
For $k=1,\ldots,G$ and $i=1,\ldots,m_k$, define
$$
\Gamma_{k,i}(\bm{\Omega}) = \frac{{\bm A}_{k,i}\bullet {\bm \Omega}} {{\bm C}_{k,i}\bullet{\bm \Omega}+1}.
$$ 
Since $\mbox{rank}(\bm{A}_{k,i})=1$ for $k=1,\ldots,G$ and $i=1,\ldots,m_k$, according to the results in Sections III-B to III-D of~\cite{MainPaper}, we have
\begin{align}
r_{\sf SBF}(\mathcal{D}) &= \min_{k=1,\ldots,G \atop i=1,\ldots,m_k} \mathbb{E}_{{\bm w}\sim\mathcal{D}}\left[ \log\left(1+\frac{{\bm w}^H{\bm A}_{k,i}{\bm w}} {{\bm C}_{k,i}\bullet {\bm \Omega}+1} \right) \right] \nonumber \\
&= \mathbb{E}_{\xi \sim p}\left[ \log\left( 1 + \xi\min_{k=1,\ldots,G \atop i=1,\ldots,m_k} \Gamma_{k,i}(\bm{\Omega}) \right)\right], \label{eq:sbf-alt}
\end{align}
where for the Gaussian SBF-AF scheme, the probability density function (PDF) of $\xi$ is given by
 \begin{equation} \label{eq:gau_dist_fn}
  p(t) = p_{\sf G}(t) = e^{-t}, \quad t \ge 0,
  \end{equation}
while for the elliptic SBF-AF scheme, the PDF of $\xi$ is given by
 \begin{equation} \label{eq:ellip_dist_fn}
  p(t) = p_{\sf E}(t) = \left( 1 - \frac{1}{r} \right) \left( 1 - \frac{t}{r} \right)^{r-2}, \quad 0 \le t \le r
  \end{equation}
with $r=\mbox{rank}(\bm{\Omega})$. Moreover, 
$$
\mathbb{E}_{{\bm w}\sim\mathcal{D}} \left[ {\bm w}^H{\bm D}_\ell {\bm w} \right] = {\bm D}_\ell \bullet {\bm \Omega}, \quad \ell=0,1,\ldots,L.
$$
Thus, by the monotonicity of the logarithm, we see that Problem (SBF) is equivalent to
\begin{align}\notag
\displaystyle \max_{{\bm \Omega} \in \mathbb{H}_+^{L^2}} \quad & \min_{k=1,\ldots,G \atop i=1,\ldots,m_k} \quad \Gamma_{k,i}(\bm{\Omega}) \\\notag
\text{subject to} \,\,\,& {\bm D}_\ell \bullet {\bm \Omega}\le \bar{P}_{\ell}, \quad \ell=0,1,\ldots,L,
\end{align}
which has exactly the same form as Problem (SDR).  This implies that every optimal solution to (SDR) is also optimal for (SBF). 
\hfill $\blacksquare$

\subsection{Proof of Theorem~\ref{thm:1}} \label{proof:thm1}
Using~\eqref{eq:gamma} and~\eqref{eq:sbf-alt}, we have
\begin{align*}
& r_{\sf SDR}- r_{\sf SBF}(\mathcal{D}) \\
&= \log\left( 1+\gamma\left(\bm{W}^\star\right) \right) - \mathbb{E}_{\xi \sim p}\left[ \log\left( 1+\xi\gamma\left( \bm{W}^\star \right) \right) \right]
\end{align*}
when $\bm{\Omega}=\bm{W}^\star$.  Now, let $g_p:{\mathbb R}_+\rightarrow{\mathbb R}$ be the function defined by
$$ g_p(y) = \log(1+y) - \mathbb{E}_{\xi \sim p}\left[ \log(1+\xi y) \right]. $$
For the Gaussian SBF-AF scheme, the PDF of $\xi$ is given by~\eqref{eq:gau_dist_fn}.  By Jensen's inequality, we have
$$
g_{p_{\sf G}}'(y) \ge \left( \frac{1}{ 1 + y } -  \frac{ \mathbb{E}_{\xi \sim p_{\sf G}} [ \xi ] }{1+y\mathbb{E}_{\xi \sim p_{\sf G}} [ \xi ] } \right) y = 0,
$$
which implies that $g$ is non-decreasing. This, together with~\cite[Theorem 1]{MainPaper}, yields
$$ r_{\sf SDR}- r_{\sf SBF}({\sf G}) \le g_{p_{\sf G}}(+\infty) = 0.5772. $$

For the elliptic SBF-AF scheme, the PDF of $\xi$ is given by~\eqref{eq:ellip_dist_fn}.  It is known that ${\mathbb E}_{\xi \sim p_{\sf E}}[\xi]=1$; see, e.g.,~\cite[Fact 3]{MainPaper}.  Hence, $g_{p_{\sf E}}$ is also non-decreasing, which implies that $r_{\sf SDR}- r_{\sf SBF}({\sf E}) \le g_{p_{\sf E}}(+\infty)$. To determine $g_{p_{\sf E}}(+\infty)$, we first use \eqref{eq:ellip_dist_fn} to compute the elliptic SBF-AF rate as shown at the top of the next page.
\begin{figure*}[!t]
\normalsize
\begin{align}
   &r_{\sf SBF}({\sf E})= \left(1-\frac{1}{r}\right)\int_0^r \log\left( 1+t\gamma\left(\bm{W}^\star\right) \right) \left(1-\frac{t}{r}\right)^{r-2} dt \label{eq:exp-def} \\
   \noalign{\medskip}
   &=\int_0^r \frac{\gamma\left(\bm{W}^\star\right)}{1+t\gamma\left(\bm{W}^\star\right)} \left( 1-\frac{t}{r} \right)^{r-1}dt \label{eq:ellip-ibp} \\
   \noalign{\medskip}
   &= \int_1^{1+r\gamma\left(\bm{W}^\star\right)} \frac{1}{y} \left( 1 - \frac{y-1}{r\gamma\left(\bm{W}^\star\right)} \right)^{r-1} dy \label{eq:ellip-cvar} \\
   \noalign{\medskip}
   &= \left( 1 + \frac{1}{r\gamma\left(\bm{W}^\star\right)} \right)^{r-1} \int_1^{1+r\gamma\left(\bm{W}^\star\right)} \frac{1}{y} \left( 1 - \frac{y}{1+r\gamma\left(\bm{W}^\star\right)} \right)^{r-1} dy \nonumber \\
   \noalign{\medskip}
   &= \left( 1 + \frac{1}{r\gamma\left(\bm{W}^\star\right)} \right)^{r-1} \int_1^{1+r\gamma\left(\bm{W}^\star\right)} \left[ \frac{1}{y} + \sum_{k=1}^{r-1} {r-1 \choose k} (-1)^k \frac{y^{k-1}}{\left( 1+r\gamma\left(\bm{W}^\star\right) \right)^k} \right] dy \label{eq:ellip-bin-thm} \\
   \noalign{\medskip}
   &= \left( 1 + \frac{1}{r\gamma\left(\bm{W}^\star\right)} \right)^{r-1} \left[ \log\left( 1+r\gamma\left(\bm{W}^\star\right) \right) + \sum_{k=1}^{r-1} {r-1 \choose k} \frac{(-1)^k}{k} \left( 1 - \frac{1}{\left( 1+r\gamma\left(\bm{W}^\star\right) \right)^k} \right) \right] \nonumber \\
   \noalign{\medskip}
   &= \left( 1 + \frac{1}{r\gamma\left(\bm{W}^\star\right)} \right)^{r-1} \left[ \log\left( 1+r\gamma\left(\bm{W}^\star\right) \right) - \sum_{k=1}^{r-1}\frac{1}{k} - \sum_{k=1}^{r-1} {r-1 \choose k} \frac{(-1)^k}{k\left( 1+r\gamma\left(\bm{W}^\star\right) \right)^k} \right]. \label{eq:ellip-har-sum}
\end{align}
\hrulefill
\vspace*{4pt}
\end{figure*}
Note that~\eqref{eq:exp-def} follows from the definition of expectation;~\eqref{eq:ellip-ibp} follows from integration by parts; \eqref{eq:ellip-cvar} follows from the change of variable $y=1+t\gamma\left(\bm{W}^\star\right)$; \eqref{eq:ellip-bin-thm} follows from the binomial theorem; \eqref{eq:ellip-har-sum} follows from the identity
\begin{equation} \label{eq:bin-sum-1}
   \sum_{k=1}^n {n \choose k} \frac{(-1)^k}{k} = -\sum_{k=1}^n \frac{1}{k}
\end{equation}
(see~\cite[Formula 0.155(4)]{bk:Gradshteyn}). Therefore, 
\begin{align*}
g_{p_{\sf E}}(y) &= \log(1+y) \\ 
&\quad- \left( 1 + \frac{1}{ry} \right)^{r-1} \Bigg[ \log(1+ry) \\
&\quad- \sum_{k=1}^{r-1}\frac{1}{k} - \sum_{k=1}^{r-1} {r-1 \choose k} \frac{(-1)^k}{k(1+ry)^k} \Bigg].  
\end{align*}
Now, by the l'H\^{o}pital rule, we have
$$ g_{p_{\sf E}}(+\infty) = \lim_{y \rightarrow \infty} g_{p_{\sf E}}(y) = \sum_{k=1}^{r-1} \frac{1}{k} - \log(r). $$
To complete the proof, we simply use the fact that the function $r \mapsto \sum_{k=1}^{r-1} \frac{1}{k} - \log(r)$ is strictly increasing and tends to $0.5772$ as $r \rightarrow \infty$ (see, e.g., \cite[Formula 0.131]{bk:Gradshteyn}).
\hfill $\blacksquare$

\end{document}